\documentclass[10pt,journal,final,twocolumn]{IEEEtran}

\IEEEoverridecommandlockouts

\usepackage{amsmath,amssymb,amsfonts}
\usepackage{cite}
\usepackage{subfigure}
\usepackage{graphicx}
\usepackage{textcomp}
\usepackage{xcolor}
\usepackage{times}
\usepackage{subfigure}         
\usepackage{amssymb,amsmath}

\usepackage{booktabs}

\usepackage{acronym}  
\usepackage{balance}

\usepackage{epstopdf}

\usepackage{bm}   
      
\usepackage{algorithm} 

\usepackage{algorithmic}

\usepackage{lipsum}
\usepackage{stfloats}
\usepackage{url}

\usepackage{setspace}

\allowdisplaybreaks[4]

\newtheorem{definition}{\bf Definition}

\newtheorem{lemma}{\bf Lemma}

\definecolor{BLUE}{rgb}{0,0,1}

\acrodef{siso}[SISO]{single-input single-output}%

\acrodef{nmse}[NMSE]{normalized mean square error}%

\acrodef{ris}[RIS]{reconfigurable intelligent surface}%

\acrodef{csi}[CSI]{channel state information}%
\acrodef{awgn}[AWGN]{additive white Gaussian noise}%

\acrodef{cs}[CS]{compressed sensing}%
\acrodef{ao}[AO]{alternaing optimization}%

\acrodef{bs}[BS]{base station}%
\acrodef{snr}[SNR]{signal-to-noise ratio}%
\acrodef{mmwave}[mmWave]{millimeter-wave}%
\acrodef{snr}[SNR]{signal-to-noise ratio}%
\acrodef{rf}[RF]{radio frequency}%

\acrodef{omp}[OMP]{orthogonal matching pursuit}%
\acrodef{ml}[ML]{maximum likelihood}%

\acrodef{mse}[MSE]{mean square error }%

\acrodef{sbr}[S-BAR]{successive Bayesian reconstructor}%

\acrodef{qidd}[QIDD]{quasi-infinite dimension disaster}%

\acrodef{as}[AS]{antenna selection}%

\acrodef{vamp}[VAMP]{vector approximate message passing}%

\acrodef{fas}[FAS]{fluid antenna system}%

\acrodef{swipt}[SWIPT]{simultaneous wireless information and power transfer}%
\acrodef{sinr}[SINR]{signal-to-interference-plus-noise ratio}%

\acrodef{rc}[RC]{reflection coefficient}%
\acrodef{mimo}[MIMO]{multiple-input multiple-output}%

\def\BibTeX{{\rm B\kern-.05em{\sc i\kern-.025em b}\kern-.08em
		T\kern-.1667em\lower.7ex\hbox{E}\kern-.125emX}}

\begin{document}
\title{ Successive Bayesian Reconstructor for Channel Estimation in Fluid Antenna Systems
}
\author{	
{{Zijian Zhang},~\IEEEmembership{Graduate Student Member,~IEEE}, {Jieao Zhu},~\IEEEmembership{Graduate Student Member,~IEEE}, \\{Linglong Dai},~\IEEEmembership{Fellow,~IEEE}, and {Robert W. Heath, Jr.},~\IEEEmembership{Fellow,~IEEE}\vspace{-1em}}
\thanks{Manuscript received XX Jan. 2024; revised XX Aug. 2024; accepted XX Dec. 2024. Date of publication XX XXX 2024; date of current version XX XXX 2024. 
This work was supported in part by the National Key Research and Development Program of China (Grant No. 2023YFB3811503), in part by the National Natural Science Foundation of China (Grant No. 62325106), and in part by the National Natural Science Foundation of China (Grant No.62031019). The conference version of this paper was presented at the IEEE WCNC'24, Dubai, United Arab Emirates, April 21–24, 2024 \cite{Zhan2404:Successive}. {\it (Corresponding author: Linglong Dai.)}}
\thanks{Zijian Zhang, Jieao Zhu, and Linglong Dai are with the Department of Electronic Engineering, Tsinghua University, Beijing 100084, China, as well as the Beijing National Research Center for Information Science and Technology (BNRist), Beijing 100084, China (e-mail: \{zhangzj20, zja21\}@mails.tsinghua.edu.cn, daill@tsinghua.edu.cn).}
\thanks{Robert W. Heath, Jr. is with the Department of Electrical and Computer Engineering, University of California, San Diego, 9500 Gilman Drive, La Jolla, CA 92093, USA (e-mail: rwheathjr@ucsd.edu).}
\thanks{Color versions of one or more figures in this article are available at https://doi.org/10.1109/TWC.2024.3515135.}
\thanks{Digital Object Identifier 10.1109/TWC.2024.3515135
}
}


\maketitle

\begin{abstract}

Fluid antenna systems (FASs) can reconfigure their antenna locations freely within a spatially continuous space. To keep favorable antenna positions, the channel state information (CSI) acquisition for FASs is essential. While some techniques have been proposed, most existing FAS channel estimators require several channel assumptions, such as slow variation and angular-domain sparsity. When these assumptions are not reasonable, the model mismatch may lead to unpredictable performance losses. In this paper, we propose the successive Bayesian reconstructor (S-BAR) as a general solution to estimate FAS channels. Unlike model-based estimators, the proposed S-BAR is prior-aided, which builds the experiential kernel for CSI acquisition. Inspired by Bayesian regression, the key idea of S-BAR is to model the FAS channels as a stochastic process, whose uncertainty can be successively eliminated by kernel-based sampling and regression. In this way, the predictive mean of the regressed stochastic process can be viewed as a Bayesian channel estimator. Simulation results verify that, in both model-mismatched and model-matched cases, the proposed S-BAR can achieve higher estimation accuracy than the existing schemes.
\end{abstract}

\section{Introduction}	

Different from the conventional \ac{mimo} system where all antennas are fabricated and fixed on an array \cite{GaoZhen'23,Kuiyu'24,WangYang'23}, \acp{fas} \cite{Wong'20'CL,wong2020fluid,wong2021fluid}, also called movable antenna (MA) systems \cite{zhu2023movable,zhu2023PO,Wenyan'23}, introduce a movable structure where a few fluid antennas can switch their locations freely within a given space. In contrast to the \ac{mimo} whose antenna spacing is usually subwavelength order of magnitude, the spacing of the available locations (referred to as "ports") for fluid antennas can approach infinitesimal \cite{zhu2023PO}. This almost continuously movable feature allows \acp{fas} to keep antennas always at favorable positions, which can fully explore the diversity and multiplexing gains of a given space \cite{zhu2023Mag}. Compared to the fixed antenna systems, \acp{fas} can maintain high transmission reliability \cite{wong2020fluid} and large-scale multiple-access \cite{wong2021fluid} while using much fewer antennas and \ac{rf} components \cite{Wong'23'CL-I}. As a result, FAS has become a promising solution to effectively reduce the hardware cost and complexity of transceivers \cite{wong2022bruce}.

\subsection{Prior Work}


The original idea of reconfiguring antenna positions dated back to 2000s, when reconfigurable antennas were proposed to balance the hardware complexity and transmission performance of \ac{mimo} systems \cite{Sayeed'07'JSTSP,Vasanthan'11'TIT,Nikhil'14'TAP}. By connecting massive antennas with a few \ac{rf} chains through a switch network, reconfigurable antennas can carefully select a subset of antennas associated with the well-performed channels for transmission \cite{Rial'16,Ahmed'16}. Thanks to the advances in the microwave field, more position-flexible fluid antennas were proposed \cite{paracha2019liquid,Fortuny'19}.  Due to their compact hardware structures based on cheap materials, the cost and complexity of fluid antennas are low. For example, in \cite{Cetiner'04}, \ac{rf} micro-electro-mechanical systems (MEMSs) were integrated into some radiation patches to reconfigure their positions. In \cite{Hayes'12}, a fluid antenna made of liquid metals could move continuously in a non-conductive tube. In \cite{Daniel'14}, pixel-like switches were utilized to design an electronically reconfigurable surface. 
By properly configuring multiple analog switches that link massive patches, the positions of radiation points can be reconfigured in several microseconds. As a result, these highly flexible structures are promising to explore the diversity and multiplexing gains of a limited space \cite{Wong'20'CL,wong2020fluid,wong2021fluid,zhu2023movable,zhu2023PO,Wenyan'23,zhu2023Mag}.


Up to now, the existing works on \acp{fas} have covered a wide range, mainly focusing on the performance analysis \cite{Wong'20'CL,wong2020fluid,wong2021fluid,wong2022closed,mukherjee2022level,tlebaldiyeva2022enhancing}, port selection \cite{chai2022port,wong2022fast,Waqar'23}, beamforming design \cite{zhu2023movable,Wenyan'23,cheng2023sum,zhu2023PO,wu2023movable,xiao2023multiuser}, and channel estimation \cite{Skouroumounis'23,ma2023compressed,Rujian'23}. For example, the ergodic capacity and the lower-bound capacity of \acp{fas} were derived in \cite{Wong'20'CL}. Then, the multiplexing gains of multi-user \acp{fas} were analyzed in \cite{wong2021fluid} and \cite{wong2023opportunistic}. By combining machine learning methods and analytical approximations, the port selection for \acp{fas} was studied in \cite{chai2022port,wong2022fast,Waqar'23} to realize fast antenna placements. To maximize the sum-rate in \acp{fas}, \ac{ao}-based algorithms that jointly optimize the antenna positions and beamformers were proposed in \cite{Wenyan'23} and \cite{cheng2023sum}. To reduce the power consumption of \acp{fas}, sequential convex
approximation (SCA)-based algorithms for \ac{fas} beamforming were proposed in \cite{zhu2023PO} and \cite{wu2023movable}.




Despite the encouraging prospects of \acp{fas}, such as high capacity and low power consumption \cite{wong2020fluid,wong2021fluid,zhu2023movable}, these expected gains are hard to achieve in practice. The transmission performance of \acp{fas} heavily relies on the positions of fluid antennas \cite{chai2022port,wong2022fast,Waqar'23}. To ensure favorable antenna placements, the \ac{csi} of available locations is essential, and most existing works have assumed the perfectly known \ac{csi} \cite{zhu2023movable,Wenyan'23,cheng2023sum,zhu2023PO,wu2023movable,xiao2023multiuser}. However, the channel estimation for \acp{fas} is challenging. Particularly, to fully obtain the performance gains provided by antenna mobility, the fluid antennas should move almost continuously in a given space \cite{wong2020fluid}. Different from conventional \ac{mimo} which has large antenna spacing, the available locations (i.e., the ports) of fluid antennas are quite dense, leading to very high-dimensional port channels. Within a coherence time, it requires a large number of pilots for \ac{csi} acquisition in \acp{fas}. Besides, limited by the hardware structure of \acp{fas}, only a few ports can be connected to \ac{rf} chains for pilot reception within the coherence time, which exacerbates the difficulty of channel estimation \cite{Skouroumounis'23,ma2023compressed,Rujian'23}. As a result, \acp{fas} can not acquire the precise \ac{csi} in real-time, which finally bottlenecks their performances in practice.




To tackle the above issue, some pilot-reduced \ac{fas} channel estimators have been proposed \cite{Skouroumounis'23,ma2023compressed,Rujian'23}. For example, in \cite{Skouroumounis'23}, a sequential linear minimum mean square error (SeLMMSE) method was proposed. In each subframe, the fluid antennas move to some equally-spaced ports for channel measurements \cite{Skouroumounis'23}. By assuming the channels are slow-varying in a short distance, the channels of those unmeasured ports are assumed to be equal to those of their nearby measured ports. In \cite{ma2023compressed}, by assuming the angular-domain sparsity of \ac{fas} channels, a \ac{cs}-based estimator was proposed. Employing \ac{omp}, this estimator can estimate the channel parameters including the angles of departure (AoDs), angles of arrival (AoAs), and gains of multiple paths. In \cite{Rujian'23}, by assuming the perfectly known AoAs of all channel paths, the gains of \ac{fas} channels can be estimated by a least-square (LS)-based method. Although these methods can achieve channel estimation for \acp{fas}, they heavily rely on some channel assumptions, such as the spatially slow variation \cite{Skouroumounis'23}, angular-domain sparsity \cite{ma2023compressed}, and known AoAs \cite{Rujian'23}. When these assumptions are not reasonable, the model mismatch may lead to unpredictable performance losses.

\subsection{Our Contributions}
In this paper, we propose a \ac{sbr} as a general solution to \ac{fas} channel estimation\footnote{Simulation codes are provided to reproduce the results in this article: http://oa.ee.tsinghua.edu.cn/dailinglong/publications/publications.html.}. Our contributions are summarized as follows.
\begin{itemize}
	\item {\bf Bayesian regression in FASs:} To the best of our knowledge, the idea of Bayesian regression \cite{srinivas2012information} is introduced into \ac{fas} channel estimation for the first time. Firstly, fluid antennas can switch their locations among ports to measure channels, resembling a sampling process. Secondly, since the ports are densely deployed, the \ac{fas} channels have strong correlation. These two properties motivate us to build a kernel that characterizes the channel correlations and then construct channels by kernel-based sampling and regression. By carefully selecting kernels and ports for channel measurements, the idea of Bayesian regression is suitable to reconstruct FAS channels in a non-parametric manner.
	\item {\bf General FAS channel estimator:} The S-BAR is proposed as a general solution to estimate \ac{fas} channels without assuming channel models. The key idea is to model the \ac{fas} channels as a stochastic process, whose uncertainty can be successively eliminated by kernel-based sampling and regression. The proposed S-BAR is realized in two stages. In the first stage, following the greedy-sampling idea of Bayesian regression, the switch matrix for FASs is designed by maximizing the mutual-information increment (MII) between two adjacent sampling points. In the second stage, according to the designed switch matrix, the channel observations of the select ports are combined with the kernels for process regression,. Finally, the regressed stochastic process can be viewed as the Bayesian estimator of \ac{fas} channels.
	\item {\bf Performance analysis and simulations:} The minimum \ac{mse} achieved by the proposed S-BAR is derived, which only depends on the eigenvalues of channel covariance and noise power. Then, numerical simulations are provided to compare the performances of different estimators. Our results show that, in both model-mismatched and model-matched cases, the proposed S-BAR outperforms the benchmark schemes with channel assumptions, and the estimation accuracy of S-BAR can increase by an order of magnitude. In addition, thanks to the hybrid offline and online implementation of S-BAR, the computational complexity of online employing S-BAR is only linear with the number of ports. 
\end{itemize}


\subsection{Organization and Notation}
\textit{Organization:} The rest of this paper is organized as follows. In Section \ref{sec:model}, the system model of an \ac{fas} is introduced, and the existing solutions of \ac{fas} channel estimation are reviewed. In Section \ref{sec:SBE}, the principle and the implementation of the proposed S-BAR are illustrated. In Section \ref{sec:PA_KS}, the performance of S-BAR is analyzed, and the kernel selection for S-BAR is discussed. In Section \ref{sec:sim}, simulations are carried out to compare the estimation performances of different schemes. Finally, in Section \ref{sec:con}, conclusions are drawn and future works are discussed.

\textit{Notation:} ${[\cdot]^{-1}}$, ${[\cdot]^{\dag}}$, ${[\cdot]^{*}}$, ${[\cdot]^{\rm T}}$, and ${[\cdot]^{\rm H}}$ denote the inverse, pseudo-inverse, conjugate, transpose, and conjugate-transpose operations, respectively; $\|\cdot\|$ denotes the $l_2$-norm of the argument; ${\bf x}(i)$ denotes the $i$-th entry of vector ${\bf x}$; ${\bf X}({i,j})$, ${\bf X}({j,:})$ and ${\bf X}({:,j})$ denote the $(i,j)$-th entry, the $j$-th row, and the $j$-th column of matrix ${\bf X}$, respectively;
$[K]$ denotes the integer set $\{1,\cdots,K\}$; ${\rm Tr}(\cdot)$ denotes the trace of its argument; ${\mathsf{P}}(\cdot|\cdot)$ is the conditional probability density function; ${\mathsf E}_{\bf x}\left(\cdot\right)$ is the expectation operator with respect to (w.r.t) the random vector ${\bf x}$; $\Re\{\cdot\}$ denotes the real part of the argument; $\ln(\cdot)$ denotes the natural logarithm of its argument; $\mathcal{C} \mathcal{N}\!\left({\bm \mu}, {\bf \Sigma } \right)$ and $\mathcal{G} \mathcal{P}\!\left({\bm \mu}, {\bf \Sigma } \right)$ respectively denote the complex Gaussian distribution and complex Gaussian process with mean ${\bm \mu}$ and covariance ${\bf \Sigma }$; ${\cal U}\left(a,b\right)$ denotes the uniform distribution between $a$ and $b$; $\mathbf{I}_{L}$ is an $L\times L$ identity matrix; ${\bf e}_L$ is a zero-one vector with its $L$-th element being one and the other elements being zero; and $\mathbf{0}_{L}$ is a zero vector or matrix with dimension $L$.

\section{System Model and Existing Methods}\label{sec:model}
\begin{figure}[!t]
	\centering
	\includegraphics[width=0.47\textwidth]{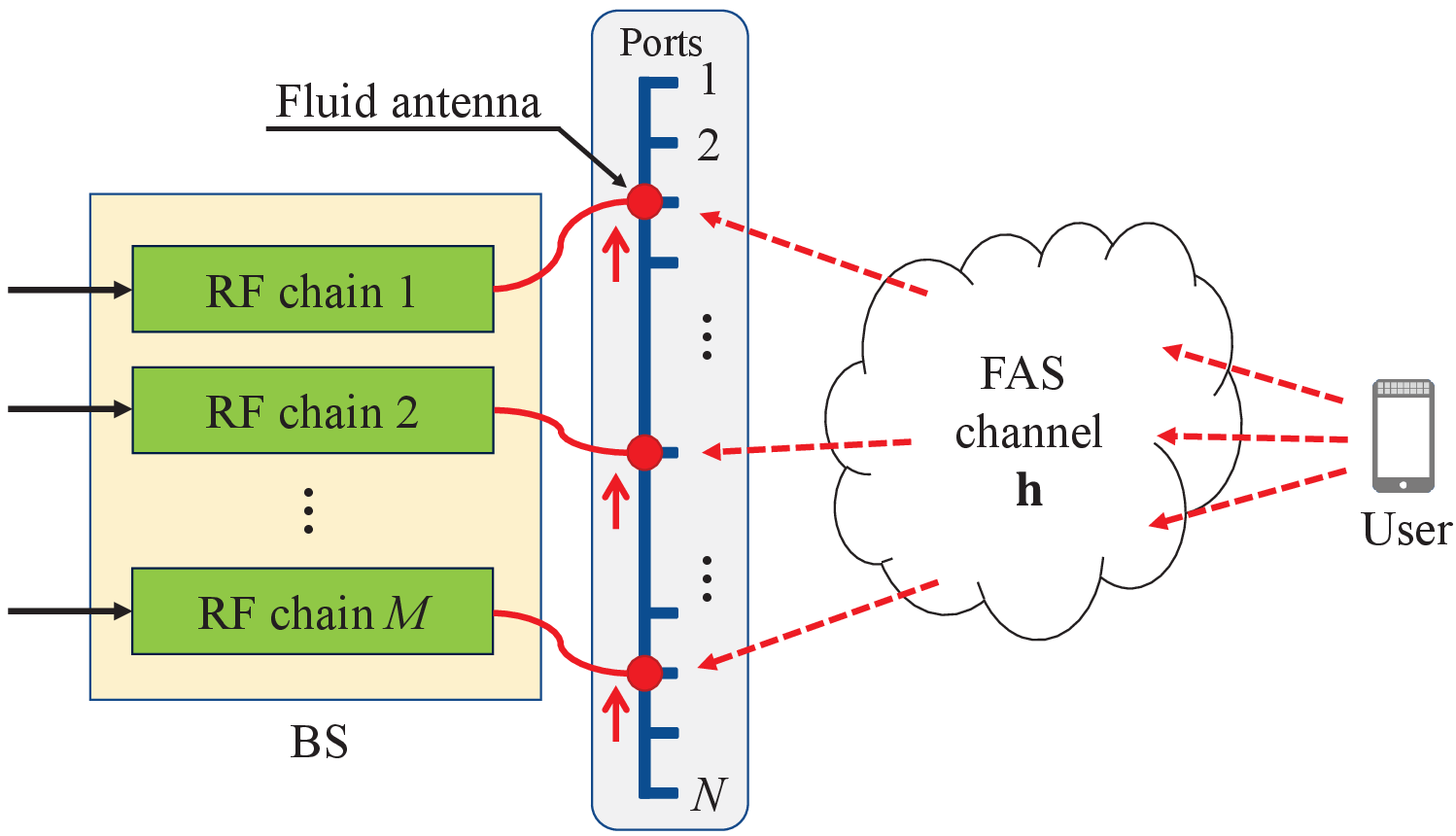}
	\caption{An illustration of channel estimation for an \ac{fas}, where one $N$-port \ac{bs} equipped with $M$ fluid antennas receives pilots from a user in the uplink.}
	\label{img:scenario}
\end{figure}
In this section, the system model is introduced in Subsection \ref{subsec:system_model}. In Subsection \ref{subsec:preliminary}, we review two typical channel estimators for \acp{fas}, including the \ac{cs}-based channel estimator and the \ac{ao}-based estimator. Finally, in Subsection II-C, the challenges and the opportunities of \ac{fas} channel estimation are discussed.
\subsection{System Model}\label{subsec:system_model}
\begin{figure}[!t]
	\centering
	\includegraphics[width=0.47\textwidth]{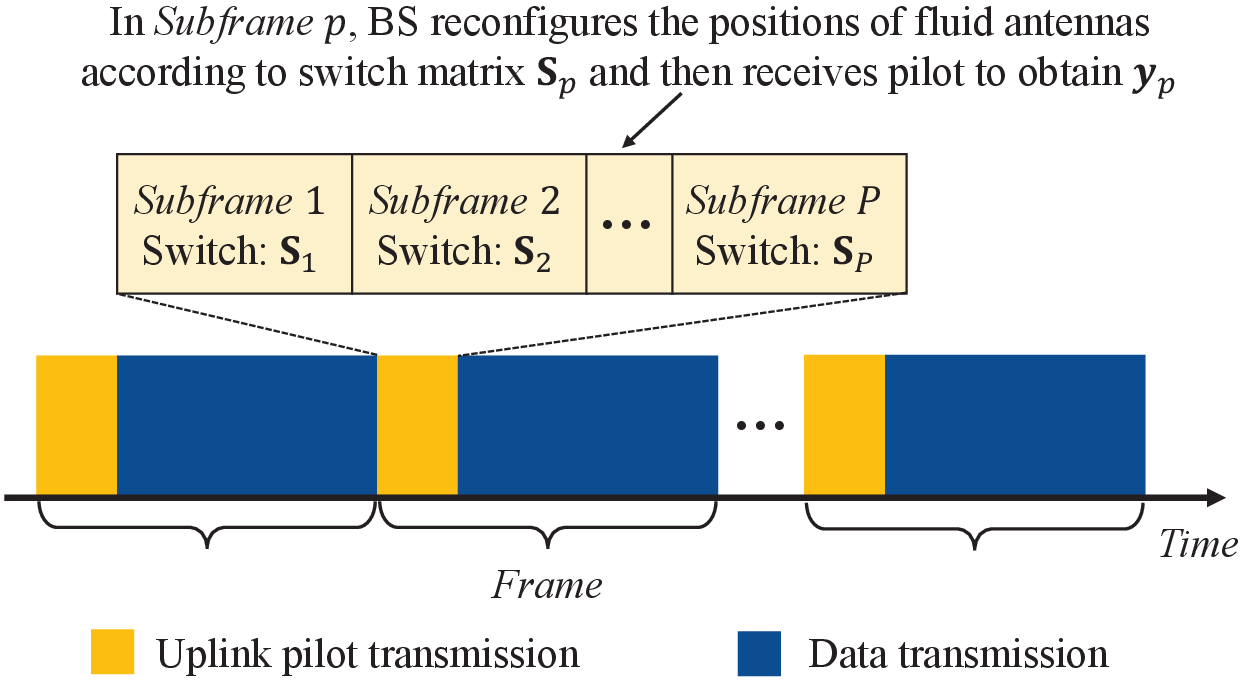}
	\caption{The frame structure for \ac{fas} channel reconstruction.}
	\label{img:frame}
\end{figure}

This paper considers the narrowband channel estimation of an uplink \ac{fas}, which consists of an $N$-port \ac{bs} equipped with $M$ fluid antennas ($M\ll N$) and a single-antenna user\footnote{ By using orthogonal pilots at different users, the proposed scheme in this paper can be  extended to the multi-user case without difficulty.
}. As shown in Fig. \ref{img:scenario}, the $N$ ports are uniformly distributed along a linear dimension at the receiver. Each fluid antenna is connected to an \ac{rf} chain for pilot reception, and the location of each antenna can be switched to one of the $N$ available port locations. Let ${\bf h}\in{\mathbb C}^N$ denote the channels of $N$ ports, and let $P$ denote the number of transmit pilots within a coherence-time frame. As shown in Fig. \ref{img:frame}, in each timeslot (subframe), fluid antennas can switch their positions to receive pilots. To characterize the locations of $M$ fluid antennas in timeslot $p$, we introduce the definition of switch matrix as follows:
\begin{definition}[Switch Matrix and Its Constraint]
Binary indicator ${\bf S}_p\in\{0,1\}^{N\times M}$ is defined as the switch matrix of multiple fluid antennas in timeslot $p$. The $(n,m)$-th entry being 1 (or 0) means that the $m$-th antenna is (or not) located at the $n$-th port. Constrained by the hardware structure, $M$ of $N$ ports should be selected in each timeslot. To satisfy this constraint, here we constrain that each column of ${\bf S}_p$ has one 1 entry, and all 1 entries in ${\bf S}_p$ are not in the same row, i.e., 
\begin{align}
\notag
{\cal S}:={\Big \{} & \|{\bf S}_p({:,m})\|=1,\|{\bf S}_p({n,:})\|\in\{0,1\}, \\ & \forall m\in[M], \forall n\in[N], \forall p\in[P]{\Big \}}.
\end{align}
Thus, we have ${\bf S}^{\rm H}_p{\bf S}_p = {\bf I}_M$ for all $p\in[P]$.
\end{definition}

Utilizing {\bf Definition 1}, the signal vector ${\bf y}_p\in{\mathbb C}^{M}$ received at the \ac{bs} in timeslot $p$ can be modeled as 
\begin{equation}\label{eqn:y_p}
{\bf y}_p = {\bf S}^{\rm H}_p{\bf h}x_p + {\bf z}_p,
\end{equation}
where $x_p$ is the pilot transmitted by the user and ${\bf z}_p \sim \mathcal{C} \mathcal{N}\!\left({\bf 0}_{M},  \sigma^2{\bf I}_{M} \right)$ is the \ac{awgn} at $M$ selected ports. Without loss of generality, we assume that $x_p = 1$ for all $p\in[P]$. Considering the total $P$ timeslots for pilot transmission, we arrive at 
\begin{equation}\label{eqn:y}
	{\bf y} = {\bf S}^{\rm H}{\bf h} + {\bf z},
\end{equation}
where ${\bf y} := \left[{\bf y}_1^{\rm H},\cdots,{\bf y}_P^{\rm H}\right]^{\rm H}$, ${\bf S} := \left[{\bf S}_1,\cdots,{\bf S}_P\right]$, and ${\bf z} := \left[{\bf z}_1^{\rm H},\cdots,{\bf z}_P^{\rm H}\right]^{\rm H}$. Our goal is to reconstruct the $N$-dimensional channel $\bf h$ according to the $PM$-dimensional noisy pilot ${\bf y}$ ($PM\ll N$). To achieve this goal, existing works usually focus on \ac{cs}-based and/or \ac{ao}-based methods, which are reviewed in the next subsection.

\subsection{Existing Methods for FAS Channel Estimation}\label{subsec:preliminary}

\subsubsection{Compressed Sensing Based Estimator}

To reconstruct the high-dimensional \ac{fas} channel $\bf h$ from the low-dimensional pilot $\bf y$, a well-known method is the \ac{cs}.
Under the assumption of spatially-sparse channels, the \ac{cs}-based channel estimator can achieve considerable performance in \ac{mimo} systems with hybrid structure \cite{Gao'16'j}. For our considered \ac{fas} channel estimation, by viewing switch matrix $\bf S$ as a virtual analog precoder, \ac{cs}-based methods can be similarly adopted. 

Specifically, let ${\bf{a}}\left( {{\theta}} \right)$ denote the steering vector as a function of incident angle ${{\theta}}$, which is defined as 
\begin{equation}\label{eqn:a}
	{\bf{a}}\left( \theta \right) = \frac{1}{{\sqrt N }}\left[ {1,{e^{{\rm j}\frac{{2\pi }}{\lambda }d\cos \left( \theta  \right)}}, \cdots ,{e^{{\rm j}\frac{{2\pi }}{\lambda }\left( {N - 1} \right)d\cos \left( \theta  \right)}}} \right]^{\rm T},
\end{equation}
where $\lambda$ is the signal wavelength and $d$ is the port spacing. Assuming that the channel is spatially sparse with $C$ clusters each contributing $R$ rays, the \ac{fas} channel $\bf h$ can be approximately modeled as 
\begin{equation}\label{eqn:h}
{\bf{h}} = \sqrt {\frac{N}{{CR}}} \sum\limits_{c = 1}^C {\sum\limits_{r = 1}^R {{g_{c,r}}{\bf{a}}\left( {{\theta _{c,r}}} \right)}},
\end{equation}
where $g_{c,r}$ and $\theta_{c,r}$ are the complex path gain and the incident angle of the $r$-th ray in the $c$-th cluster, respectively. Let ${\bf F}\in{\mathbb C}^{N\times N}$ denote the discrete Fourier transform (DFT) matrix, thus $\bf h$ can be transformed into its angular-domain representation $\tilde{\bf h}$. Under the assumption of spatial sparsity, the angular-domain channel $\tilde{\bf h}$ only takes $L$ significant values in a few of its entries ($L\ll N$). In this context, the received signal $\bf y$ in (\ref{eqn:y}) can be rewritten as
\begin{equation}\label{eqn:y_a}
	{\bf y} = {\bf S}^{\rm H}{\bf F}{\tilde{\bf h}} + {\bf z} = {\bm \Psi}{\tilde{\bf h}} + {\bf z},
\end{equation}
where ${\bm \Psi} = {\bf S}^{\rm H}{\bf F}$ is the sensing matrix.  Under the principle of uniform sampling, we can assume that the elements in ${\bf S}$ are randomly selected from $\{0,1\}$ subject to ${\bf S}\in{\cal S}$. 

On can find that, (\ref{eqn:y_a}) is a standard observation equation for sparse signal reconstruction, which can be addressed by the some existing \ac{cs}-based algorithms. For example, in \cite{ma2023compressed}, an \ac{fas}-\ac{omp} reconstructor is proposed to estimate the channel parameters in \acp{fas}. After iteratively selecting $L$ columns of sensing matrix ${\bm \Psi}$ for sparse representation, the parameters of $\bf h$ including path gains and incident angles can be estimated. By substituting the estimated channel parameters into (\ref{eqn:h}), the \ac{fas}-\ac{omp} reconstructor in \cite{ma2023compressed} can be modified to acquire channel $\bf h$ explicitly, as summarized in {\bf Algorithm \ref{alg:OMP}}.

\begin{algorithm}[!t] 
	\caption{\ac{fas}-\ac{omp} reconstructor} 
	\begin{algorithmic}[1] 
		\label{alg:OMP}
		\REQUIRE  
		Number of pilots $P$, spatial sparsity $L$.
		\ENSURE  
		Reconstructed \ac{fas} channel $\hat{\bf h}$.	
		\STATE Employ randomly generated ${\bf S}\in{\cal S}$ at the \ac{bs}
		\STATE Initialization: ${\bf r} = {\bf y}$, $\varUpsilon = \varnothing$, $\bar{\bf h} = {\bf 0}_N$
		\FOR {$l \in \{1,\cdots,L\}$}
		\STATE Update correlation matrix: ${\bm \Gamma} = {\bm \Psi}^{\rm H}{\bf r}$
		\STATE Find new support: $\upsilon^\star = \mathop{\rm argmax}\limits_\upsilon|{\bm \Gamma}(\upsilon)|$
		\STATE Update support set: $\varUpsilon = \varUpsilon \cup \{\upsilon^\star\}$
		\STATE Orthogonal projection: $\bar{\bf h}\left(\varUpsilon\right) = {\bm \Psi}\left(:,\varUpsilon\right)^{\dag}{\bf y}$
		\STATE Update residual: ${\bf r} = {\bf r} - {\bm \Psi}\left(:,\varUpsilon\right)\bar{\bf h}\left(\varUpsilon\right)$
		\ENDFOR	
		\RETURN Reconstructed channel $\hat{\bf h} = {\bf F}\left(:,\varUpsilon\right)\bar{\bf h}\left(\varUpsilon\right)$ 
	\end{algorithmic}
\end{algorithm}



\subsubsection{Alternating Optimization Based Estimator}

The \ac{cs}-based channel estimators usually operate based on a gridded codebook, e.g., the DFT matrix $\bf F$. When the real channel parameter values fall between the grid points, the non-ideal grid sampling may introduce estimation error \cite{wang2012generalized}. To improve the estimation accuracy, the channel parameters can be further optimized to cope with the errors introduced by on-grid sampling \cite{Cui'22'TCOM}. Given the spatial sparsity $L$, this goal can be achieved by finding the path gains ${\bf g}:=\left[g_1,\cdots,g_L\right]^{\rm T}$ and incident angles ${\bm \theta}:=\left[\theta_1,\cdots,\theta_L\right]^{\rm T}$ that maximize the likelihood function ${\mathsf P}\left({\bf y}|{\bf g}, {\bm \theta}\right)$, i.e., the \ac{ml} estimator. Following this idea, an \ac{ao}-based method called \ac{fas}-\ac{ml} reconstructor is summarized in {\bf Algorithm \ref{alg:ML}}, which is explained as follows.

\begin{algorithm}[!t] 
	\caption{\ac{fas}-\ac{ml} reconstructor} 
	\begin{algorithmic}[1] 
		\label{alg:ML}
		\REQUIRE  
		Number of pilots $P$, spatial sparsity $L$.
		\ENSURE  
		Reconstructed \ac{fas} channel $\hat{\bf h}$.	
		\STATE Employ randomly generated ${\bf S}\in{\cal S}$ at the \ac{bs}
		\STATE Initialize $\hat {\bf{g}}$ and $\hat {\bm{\theta }}$ via a \ac{cs}-based algorithm, e.g., \ac{omp}
		\STATE Initialize step length by $\zeta = 1$
		\WHILE {no convergence of ${\mathsf P}({\bf y}|\hat{\bf g}, \hat{\bm \theta})$}
		\STATE Update path gain: $\hat{\bf g} =  {\bf B}^{\dag}(\hat{\bm{\theta }}){\bf y}$
		\STATE Update angle: $\hat{\bm{\theta }} = \hat{\bm{\theta }} - \zeta \Re\left\{ {\frac{{\partial \left\|{\bf y} - {\bf{B}}\left( {\bm{\theta }} \right){\bf{g}} \right\|^2 }}{{\partial {\bm{\theta}}}}} |_{ {\bf{g}} = \hat{\bf{g}},{\bm{\theta}}=\hat{\bm{\theta}}} \right\}$
		\STATE Update step length: $\zeta = \zeta/2$
		\ENDWHILE
		\RETURN Reconstructed channel $\hat{\bf h}= \sqrt{\frac{N}{L}}\sum\nolimits_{l = 1}^L {\hat{g_l}{\bf{a}}( {\hat{\theta _l}} )}$ 
	\end{algorithmic}
\end{algorithm}

Assume that channel $\bf h$ has the spatial sparsity as shown in (\ref{eqn:h}). For the considered problem of channel estimation, the logarithm likelihood function can be written as
\begin{equation}\label{eqn:LLF}
	\begin{aligned}
		\ln\left({\mathsf P}\left({\bf y}|{\bf g}, {\bm \theta}\right)\right) = -N\ln\left(\sigma^{2}\pi\right) - \frac{1}{\sigma^2}\left\|{\bf y} - {\bf{B}}\left( {\bm{\theta }} \right){\bf{g}} \right\|^2,
	\end{aligned}
\end{equation}
wherein ${\bf B}\left( {\bm{\theta }} \right) := \sqrt {\frac{N}{L}} {\bf{S}}^{\rm H}\left[ {{\bf{a}}\left( {{\theta _1}} \right), \cdots ,{\bf{a}}\left( {{\theta _L}} \right)} \right]$. Then, an \ac{ml}-based estimation can be achieved by solving 
\begin{equation}
\{{{\bf{g}}^\star, {\bm{\theta }}^\star} \} = \mathop{\arg\!\min}\limits_{{\bf{g}},{\bm{\theta }}} \left\|{\bf y} - {\bf{B}}\left( {\bm{\theta }} \right){\bf{g}} \right\|^2.
\end{equation}
However, due to the coupled variables ${\bf g}$ and ${\bm \theta}$, the optimal ${\bf{g}}$ and ${\bm{\theta }}$ are hard to be obtained simultaneously. As a compromise,  ${\bf{g}}$ and ${\bm{\theta }}$ are usually updated in an alternating way to approach the sub-optimal solution \cite{admm}. By fixing one and optimize the other variables, the likelihood function ${\mathsf P}\left({\bf y}|{\bf g}, {\bm \theta}\right)$ can be maximized through an \ac{ao} process. Finally, the \ac{fas}-\ac{ml} reconstructor in {\bf Algorithm \ref{alg:ML}} can be obtained. After sufficient iterations, all variable updates make ${\mathsf P}\left({\bf y}|{\bf g}, {\bm \theta}\right)$ monotonically increase, thus {\bf Algorithm \ref{alg:ML}} is always convergeble.

\subsection{Challenge and Opportunity}
{\it Challenge:} Despite the feasibility of existing \ac{fas} channel estimators, two inherent drawbacks have bottlenecked their estimation accuracy. First, most existing estimators are parametric algorithms, which usually follow the assumption of spatially sparse channels \cite{ma2023compressed}. In practical scenarios, the model mismatch may lead to unpredictable performance loss. Second, most existing estimators are based on the randomly generated zero-one distributed $\bf S$, i.e., the switch matrix. Different from the \ac{mimo} analog precoder whose elements are all unit-modulus \cite{Gao'16'j}, only a few elements in $\bf S$ are one while the others are all zero. It implies that, the received pilot $\bf y$ only contains the information of a few channels associated with the selected ports. Consequently, high pilot overhead (i.e., a large $P$) may be required to accurately reconstruct channel $\bf h$.



{\it Opportunity:} Different from \ac{mimo} that measures all channels simultaneously, fluid antennas switch their locations among the ports to measure channels in a successive way, which resembles a sampling process. Besides, unlike the conventional \ac{mimo} whose typical antenna spacing is usually $\lambda/2$, the port spacing of \acp{fas} $d$ is usually smaller, such as $\lambda/10$ \cite{Wong'20'CL,wong2020fluid,wong2021fluid}. Thus, the channels of several ports closer to each other are strongly correlated \cite{Jiancheng'23}. It suggests that, when the channel of a port is sampled (measured), the uncertainty of the channels associated with its close-by ports can be partially eliminated. These properties inspire that, by carefully designing a sampling sequence and conducting regression, the \ac{fas} channels can be reconstructed in a non-parametric way. 



\section{Proposed Successive Bayesian Reconstructor}\label{sec:SBE}

By building experiential kernels for an objective function, Bayesian regression can determine the sampling strategy and reconstruct the objective functions via kernel-based regression. Thanks to the sampler-like fluid antennas and strongly correlated channels in \acp{fas}, Bayesian regression perfectly matches the problem of \ac{fas} channel estimation. 

Following this idea, in this section, we propose the \ac{sbr} as a high-accuracy solution to \ac{fas} channel estimation \cite{srinivas2012information}. Firstly, the Bayesian regression is introduced in Subsection \ref{subsec:BLR}. Then, in Subsection \ref{subsec:S-BAR}, the proposed S-BAR is illustrated. Finally, in Subsection \ref{subsec:two_stage}, the practical implementation of S-BAR is discussed.

\subsection{Bayesian Linear Regression}\label{subsec:BLR}

Without making any prior assumptions, the attempt to recover the function $f({\bf x})$ from a limited number of samples appears to be a challenging endeavor. Modeling $f({\bf x})$ as a realization of a stochastic process offers an elegant means of specifying function properties in a non-parametric manner. Under this framework, Bayesian linear regression, also called Gaussian process regression (GPR) or Kriging method \cite{oliver1990kriging}, has become a popular solution. Specifically, function $f({\bf x})$ can be modeled as a sample of Gaussian process ${\mathcal{GP}}\left(\mu\left({\bf x}\right), k\left({\bf x},{\bf x}'\right)\right)$, where any finite subset follows a consistent multivariate Gaussian distribution. It is completely specified by its mean function $\mu\left({\bf x}\right)$, which can be assumed to be zero, and its kernel function $k\left({\bf x},{\bf x}'\right)$, which encodes smoothness properties of recovered $f({\bf x})$. For clarity, here we summarize the Bayesian linear regression in {\bf Algorithm \ref{alg:blr}}, and the detailed explanations are provided as follows.

\begin{algorithm}[!t] 
	\caption{Bayesian Linear Regression} 
	\begin{algorithmic}[1]
		\label{alg:blr}
		\REQUIRE  
		Definition domain $\cal S$, prior ${\mathcal{GP}}\left(\mu\left({\bf x}\right), k\left({\bf x},{\bf x}'\right)\right)$, tolerance threshold $\varepsilon$.
		\ENSURE 
		Reconstructed function $\hat{f}({\bf x})$.	
		\STATE Initialization: $t = 0$, ${\cal A}^0= \varnothing$, $k^0\left({\bf x},{\bf x}'\right) = k\left({\bf x},{\bf x}'\right)$
		\WHILE{${k^t\left({\bf x},{\bf x}\right)} > \varepsilon$ for an ${\bf x}\in{\cal S}$}
		\STATE Timeslot update: $t = t+1$
		\STATE Sample selection: ${\bf x}^{t} = \mathop {\arg \max }\limits_{{\bf x}\in {\cal S}/{\cal A}^{t-1}} ~ {k^{t-1}\left({\bf x},{\bf x}\right)}$
		\STATE Update measured points: ${\cal A}^{t} = {\cal A}^{t-1} \cup \{{\bf x}^{t}\}$
		\STATE Measurement: $\gamma^t = f({\bf x}^t) + n_t$
		\STATE Posterior update: Calculate mean ${\mu^t\left({\bf x}\right)}$ and covariance ${k^t\left({\bf x},{\bf x}'\right)}$ by (\ref{eqn:mu_x}) and (\ref{eqn:k_x_x'}) for all ${\bf x},{\bf x}'\in{\cal S}$, respectively
		\ENDWHILE
		\STATE Reconstruction: $\hat{f}({\bf x}) = {\mu^t\left({\bf x}\right)}$
		\RETURN Reconstructed function $\hat{f}({\bf x})$ 
	\end{algorithmic}
\end{algorithm}

In timeslot $t$, consider a prior ${\mathcal{GP}}\left(\mu\left({\bf x}\right), k\left({\bf x},{\bf x}'\right)\right)$ over $f({\bf x})$. Let ${\bm \gamma}^t := [\gamma^1,\cdots,\gamma^t]^{\rm T}$ denote $t$ noisy measurements for points in ${\cal A}^t := \{{\bf x}^1,\cdots,{\bf x}^t\}$, where $\gamma^i = f({\bf x}^i) + n_i$ with $n_i\sim{\cal{CN}}\left(0, \delta^2\right)$. The joint probability distribution of $f({\bf x})$ and ${\bm \gamma}^t$ satisfies
\begin{equation}
	\begin{aligned}
		\left[ \begin{array}{l}
			f\left( {\bf{x}} \right)\\
			{{\bm{\gamma }}^t}
		\end{array} \right]
		\sim 
		{\cal CN}\!
		\left( 
		{
			\left[ \begin{array}{c}
				{\mu\left({\bf x}\right)}\\
				{{\bm \mu}^t}
			\end{array} \right],
			\left[ {\begin{array}{*{5}{c}}
					{k\left({\bf x},{\bf x}\right)}&({\bf k}^t({\bf x}))^{\rm H}\\
					{\bf k}^t({\bf x})&{{\bf K}^t + \delta^2{\bf I}_t}
			\end{array}} \right]
		} 
		\right),
	\end{aligned}
\end{equation}
where ${\bf k}^t({\bf x}): = \left[k\left({\bf x}^1,{\bf x}\right),\cdots,k\left({\bf x}^t,{\bf x}\right)\right]^{\rm T}$; ${\bm \mu}^t: = \left[\mu\left({\bf x}^1\right),\cdots,\mu\left({\bf x}^t\right)\right]^{\rm T}$
; and the $(i,j)$-th entry of ${\bf K}^t\in{\mathbb C}^{t\times t}$ is $k\left({\bf x}^i,{\bf x}^j\right)$, for all $i,j\in\{1,\cdots,t\}$. It is easy to prove that, given ${\bm{\gamma }}^t$, the posterior over $f({\bf x})$ is also a Gaussian process, with its mean and covariance being:
\begin{align}
\label{eqn:mu_x}
{\mu^t\left({\bf x}\right)} &= \mu\left({\bf x}\right) + \left({\bf k}^t({\bf x})\right)^{\rm H}\! \left( {\bf K}^t \!+\! \delta^2{\bf I}_t \right)^{\!-1}\!\left({\bm \gamma}^t-{\bm \mu}^t\right), \\ 
\label{eqn:k_x_x'}
{k^t\left({\bf x},{\bf x}'\right)} &= k\left({\bf x},{\bf x}'\right) \!-\! \left({\bf k}^t({\bf x})\right)^{\rm H} \! \left( {\bf K}^t + \delta^2{\bf I}_t \right)^{-1}\!{\bf k}^t({\bf x}').
\end{align}

Then, the next candidate point to be sampled, i.e., ${\bf x}^{t+1}$, can be determined based on the updated posterior. From the information perspective, the points to be sampled should be as uncorrelated as possible. In the case of one-by-one sampling, sampling the point with the maximum posterior variance can obtain the most information. Therefore, by assuming that ${\bf x}\in {\cal S}$, ${\bf x}^{t+1}$ can be chosen according to
\begin{align}
\label{eqn:x_t+1}
{\bf x}^{t+1} = \mathop {\arg\!\max }\limits_{{\bf x}\in {\cal S}/{\cal A}^{t}} ~ {k^{t}\left({\bf x},{\bf x}\right)},
\end{align}
where $/$ is the set difference. By updating ${\cal A}^{t}$ and ${\bm \gamma}^{t}$ accordingly, the variance function $k^t\left({\bf x},{\bf x}\right)$ decreases asymptotically, which means that the uncertainty of $f({\bf x})$ is gradually eliminated. After reaching the tolerance threshold $\varepsilon$, the posterior mean ${\mu^t\left({\bf x}\right)}$ in (\ref{eqn:mu_x}) can be viewed as the Bayesian estimator of $f({\bf x})$, which completes the algorithm.

\subsection{Working Principle of the Proposed S-BAR}\label{subsec:S-BAR}
In each pilot timeslot, $M$ fluid antennas can move positions and measure channels, thus the channel estimation of \acp{fas} is similar to a successive sampling process. Since the port spacing is very short, the \ac{fas} channels are strongly correlated. These features inspire us to recover $\bf h$ through optimizing the reconstructor $\hat{\bf h}$ and the switch matrix $\bf S$ jointly, which can be formulated as
\begin{align}\label{eq:problem_Bayesian}
	\min_{\hat{\bf h},{\bf S}\in{\cal S}}~{\mathsf E}_{{\bf h},{\bf z}}\left(\|{\bf h} - \hat{\bf h}\|^2\right), 
\end{align}
It is obvious that (\ref{eq:problem_Bayesian}) is a Bayesian estimation problem. To efficiently solve this problem, one can model $\bf h$ as a sample of Gaussian process ${\mathcal{GP}}\left({\bf 0}_N, {\bm \Sigma}\right)$. In particular, semidefinite Hermitian matrix ${\bm \Sigma}\in{\mathbb C}^{N\times N}$ is called the kernel, which characterizes (but does not need to be) the prior covariance of $\bf h$. Given the distribution of channels and noises, the joint probability distribution of $\bf h$ and $\bf y$ satisfies
\begin{equation}
	\begin{aligned}
		\left[ \begin{array}{l}
			{\bf{h}}\\
			{{\bf{y}}}
		\end{array} \right] \sim \mathcal{CN} \left(
		\left[ \begin{array}{c}
			{\bf 0}_{N} \\
			{\bf 0}_{PM}
		\end{array} \right], 
		\left[ 
		{\begin{array}{*{20}{c}}
				{{{\bf{\Sigma }}}}&{{\bf{\Sigma 
					}}}{{\bf{S}}}\\{{\bf{S}}^{\rm H}}
				{{\bf{\Sigma 
					}}}& {\bf S}^{\rm H}{\bf{\Sigma 
				}}{\bf S} + \sigma^2{\bf I}_{PM}
		\end{array}} \right]\right),
	\end{aligned}
\end{equation}
Therefore, for given ${\bf{y}}$, the posterior mean and the posterior covariance of $\bf h$ can be calculated by:
\begin{align}
	\label{eqn:mu_O}
	{{\bm{\mu }}_{{\bf{h}}|{\bf{y}}}} &= {\bf{\Sigma S}}{\left( {{{\bf{S}}^{\rm{H}}}{\bf{\Sigma S}} + {\sigma ^2}{{\bf{I}}_{PM}}} \right)^{ - 1}}{\bf{y}}, \\
	\label{eqn:Sigma_O}
	{{\bf{\Sigma }}_{{\bf{h}}|{\bf{y}}}} &= {\bf{\Sigma }} - {\bf{\Sigma S}}{\left( {{{\bf{S}}^{\rm{H}}}{\bf{\Sigma S}} + {\sigma ^2}{{\bf{I}}_{PM}}} \right)^{ - 1}}{{\bf{S}}^{\rm{H}}}{\bf{\Sigma }}.
\end{align}
Given a switch matrix $\bf S$, it is straightforward to prove that the optimal reconstructor $\hat {\bf h}$ is exactly the posterior mean, i.e., $\hat{\bf h}={{\bm{\mu }}_{{\bf{h}}|{\bf{y}}}}$, and the \ac{mse} in (\ref{eq:problem_Bayesian}) can be rewritten as the trace of posterior covariance, i.e., ${\mathsf E}_{{\bf h},{\bf z}}(\|{\bf h} - \hat{\bf h}\|^2)={\rm Tr}( {{\bf{\Sigma }}_{{\bf{h}}|{\bf{y}}}})$. Thus, problem (\ref{eq:problem_Bayesian}) can be solved by finding a proper switch matrix ${\bf S}$ that minimizes ${\rm Tr}( {{\bf{\Sigma }}_{{\bf{h}}|{\bf{y}}}})$, i.e.,
\begin{align}\label{eq:refor_problem}
{\bf{S}} = {\mathop{\arg\!\min}_{\bf{S}\in{\cal S}}}~{\rm{Tr}}\left( {{{\bf{\Sigma }}_{{\bf{h}}|{\bf{y}}}}} \right).
\end{align}
Unfortunately, due to the constraint ${\bf S}\in{\cal S}$ and the non-convex objective ${\rm Tr}( {{\bf{\Sigma }}_{{\bf{h}}|{\bf{y}}}})$, problem (\ref{eq:refor_problem}) is still hard to solve. 

To overcome the above challenge, we adopt the greedy-sampling idea from Bayesian linear regression to design $\bf S$. The core technique is to greedily optimize $\bf S$ in a column-by-column manner by maximizing the mutual-information increment (MII) between channels and received pilots. Specifically, let ${\bf S}_t:=[{\bf s}_1,\cdots,{\bf s}_t]\in\{0,1\}^{N \times t}$ denote the first-$t$ columns of ${\bf S}$ where $1\le t\le PM$. Accordingly, the first-$t$ received signals can be written as ${\bf y}_t = {\bf S}_t^{\rm H}{\bf h} + {\bf z}_t$, wherein ${\bf z}_t$ is the first-$t$ \ac{awgn}. Given ${\bf S}_t$, we aim to design the $(t+1)$-th column ${\bf s}_{t+1}$ by maximizing the MII between received pilots and channels from $t$ to $t+1$, which can be formulated as
\begin{align}\label{eqn:MII_max_problem}
\mathop{\max}_{{\bf s}_{t+1}\in{\cal S}}~I({\bf y}_{t+1};{\bf h}) - I({\bf y}_{t};{\bf h}),
\end{align}
where the mutual information $I({\bf y}_{t};{\bf h})$ is given by
\begin{align}\label{eq:MI_y_t}
I({{\bf{y}}_t};{\bf{h}}) = \log_2 \det \left( {{\bf{I}}_t + \frac{1}{{{\sigma ^2}}}{\bf{S}}_t^{\rm{H}}{\bf{\Sigma }}{{\bf{S}}_t}} \right).
\end{align}
Furthermore, we introduce the following lemma to simplify the problem formulation:
\begin{lemma}[Equivalent problem of MII maximization]
Given ${\bf S}_t$, the MII maximization problem in  (\ref{eqn:MII_max_problem}) can be equivalently rewritten as
\begin{align}\label{eq:s_t+1_optimize}
{{\bf{s}}_{t + 1}} = \mathop {\arg\!\max }\limits_{{\bf{s}} \in {\cal S} } \;{{\bf{s}}^{\rm{H}}}{{\bf{\Sigma }}_t}{\bf{s}},
\end{align}
where ${\bf{\Sigma }}_t$ is written as
\begin{align}\label{eq:Sigma_t}
{{\bf{\Sigma }}_t} = {\bf{\Sigma }} - {\bf{\Sigma }}{{\bf{S}}_t}{\left( {{\bf{S}}_t^{\rm{H}}{\bf{\Sigma }}{{\bf{S}}_t} + {\sigma ^2}{{\bf{I}}_t}} \right)^{ - 1}}{\bf{S}}_t^{\rm{H}}{\bf{\Sigma }},
\end{align}
which is exactly the posterior covariance of $\bf h$ for given ${\bf y}_t$. In particular, we have ${\bf \Sigma}_0={\bf \Sigma}$ and ${\bf \Sigma}_{PM}={\bf \Sigma}_{{\bf h}|{\bf y}}$.
\end{lemma}
\begin{IEEEproof}
Constructive proof is given in Appendix A.	
\end{IEEEproof}
Recalling the hardware constraint of \acp{fas} $\cal S$, the zero-one switch vector ${\bf s}_{t+1}$ should only have one 1 entry, such that the corresponding fluid antenna is positioned at one port. Thus, one may feel that problem (\ref{eq:s_t+1_optimize}) can be optimally solved by finding the sampling index $n_{t+1} = \mathop {\arg \max }\nolimits_{n\in [N]} ~ {{\bm \Sigma }_{t}(n,n)}$ and then setting ${\bf s}_{t+1}={\bf e}_{n_{t+1}}$. However, in fact, selecting index from the full port set $[N]$ is not correct, because multiple antennas cannot be located at the same port in a single timeslot. That is to say, all 1 entries in switch matrix ${\bf S}_p$ cannot be in the same row, as illustrated in {\bf Definition 1}. 

To address the port conflict issue while optimizing ${{\bf{s}}_{t + 1}}$, $\Omega_p$ is introduced to represent the sampling sequence, which is composed of the indexes of the selected ports in the $p$-th timeslot. $\Omega_p$ is initialized as an empty set. Once a port is selected, it will be added to $\Omega_p$, and then the next sampling index can be selected from the difference set  $[N]/\Omega_p$. Thus, the optimal solution to problem (\ref{eq:s_t+1_optimize}) can be obtained by
\begin{subequations}\label{eq:n_t+1}
\begin{align}
	n_{t+1} &= \mathop {\arg\!\max }\limits_{n\in [N]/\Omega_p} ~ {{\bm \Sigma }_{t}(n,n)}, \\
	{{\bf{s }}_{t+1}} & = {\bf e}_{n_{t+1}}.
\end{align}
\end{subequations}
In this way, constraint $\cal S$ can be naturally satisfied during the design process. This ensures that the designed $\bf S$ is practically implementable in \ac{fas} systems. After determining $n_{t+1}$, we can update $\Omega_p$ by $\Omega_p \cup \{n_{t+1}\}$ and repeat the above process. When the number of indexes in $\Omega_p$ is equal to the number of fluid antennas $M$, we can update $\Omega_p$ in (\ref{eq:n_t+1}) by $\Omega_{p+1}$ to focus on the port selection in the $(p+1)$-th timeslot. After obtaining all $PM$ columns of switch matrix ${\bf S}:=\left[{\bf s}_1,\cdots,{\bf s}_{PM}\right]$, one can obtain the channel reconstructor by
\begin{align}\label{eq:channel_reconstrcutor}
\hat{\bf h} &= {\bf{\Sigma S}}{\left( {{{\bf{S}}^{\rm{H}}}{\bf{\Sigma S}} + {\sigma ^2}{{\bf{I}}_{PM}}} \right)^{ - 1}}{\bf{y}},
\end{align} 
which completes the algorithm. 

For better understanding, the proposed S-BAR is summarized in {\bf Algorithm \ref{alg:SBAR}}, which is realized in a hybrid offline and online manner. More details about its two-stage implementation will be discussed in the next subsection.

\begin{algorithm}[!t] 
	\caption{Proposed Successive Bayesian Reconstructor} 
	\begin{algorithmic}[1]
		\label{alg:SBAR}
		\REQUIRE  
		Number of pilots $P$, kernel ${\bm \Sigma}$.
		\ENSURE 
		Reconstructed \ac{fas} channel $\hat{\bf h}$. 
		\STATE {\it \# Stage 1 (Offline Design):}
 		\STATE Initialization: $t=0$, ${\bm \Sigma }_0={\bm \Sigma }$
		\FOR{$p\in\{1,\cdots,P\}$}
			\STATE Initialize sampling sequence at pilot $p$: $\Omega_p=\varnothing$
			\FOR{$m\in\{1,\cdots,M\}$}
			\STATE Posterior covariance update: Calculate ${\bm \Sigma }_{t}$ by (\ref{eq:Sigma_t_recursion})
			\STATE Candidate selection: $n_{t+1} = \mathop {\arg \max }\limits_{n\in [N]/\Omega_p} ~ {{\bm \Sigma }_{t}(n,n)}$
			\STATE Switch vector design: ${{\bf{s }}_{t+1}} = {\bf e}_{n_{t+1}}$
			\STATE The $p$-th sequence update: $\Omega_p = \Omega_p \cup \{n_{t+1}\}$
			\STATE Counter update: $t = t+1$
			\ENDFOR
		\ENDFOR
		\STATE Merge switch vectors: ${\bf S} = \left[{\bf s}_1,\cdots,{\bf s}_{PM}\right]$
		\STATE Weight calculation: ${\bf W} = ( {\bf S}^{\rm H}{\bm \Sigma}{\bf S} + \sigma^2{\bf I}_{PM} )^{-1} {\bf S}^{\rm H}{\bm \Sigma}$
		\STATE {\it \# Stage 2 (Online Regression):}
		\STATE Employ the designed switch matrix ${\bf S}$ at the \ac{bs}, and then obtain the received pilot: ${\bf y} = {\bf S}^{\rm H}{\bf h} + {\bf z}$
		\STATE Channel reconstruction: $\hat{\bf h} = {\bf W}^{\rm H}{\bf y}$
		\RETURN Reconstructed \ac{fas} channel $\hat{\bf h}$ 
	\end{algorithmic}
\end{algorithm}


\subsection{Hybrid Offline and Online Implementation of S-BAR}\label{subsec:two_stage}
In this subsection, the practical implementation of the proposed S-BAR is discussed. Firstly, from the derivations in the above subsection, we obtain the following three observations.
\begin{itemize}
	\item Equation (\ref{eq:channel_reconstrcutor}) indicates that, given $\bf S$, the channel reconstructor $\hat{\bf h}$ is the linear weighted sum of the received pilots ${\bf y}$, i.e., $\hat{\bf h}={\bf W}^{\rm H}{\bf y}$, where the weight ${\bf W}:={\left( {{{\bf{S}}^{\rm{H}}}{\bf{\Sigma S}} + {\sigma ^2}{{\bf{I}}_{PM}}} \right)^{ - 1}}{{\bf{S}}^{\rm{H}}}{\bf{\Sigma }}$ only relies on the kernel ${\bm \Sigma}$.
	\item Equation (\ref{eq:Sigma_t}) shows that, posterior covariance ${\bm \Sigma }_{t}$ only relies on kernel ${\bm \Sigma}$, while it is unrelated to the received pilot ${\bf y}$ in instantaneous channel estimation.
    \item Equation (\ref{eq:n_t+1}) suggests that the \ac{fas} port selection only relies on the posterior covariance ${\bm \Sigma }_{t}$. 
\end{itemize}
These observations reveal that, the switch matrix $\bf S$ and the weight $\bf W$ for reconstructing $\bf h$ can be designed offline and then deployed online for regression. Therefore, the implementation of the proposed S-BAR can be realized in the two stages as shown in {\bf Algorithm 4}, so that the complexity of employing the S-BAR online can be significantly reduced. The details are explained as follows.

\subsubsection{Stage 1 (Offline Design)}
Since switch matrix is determined by the posterior covariance ${\bm \Sigma }_{t}$, while ${\bm \Sigma }_{t}$ only relies on the kernel ${\bm \Sigma}$. ${\bf S}\in{\{0,1\}}^{N \times PM}$ and the weight ${\bf W}\in{\mathbb C}^{PM}$ for reconstructing ${\bf h}\in{\mathbb C}^N$ can be designed offline at the first stage. By updating ${\bm \Sigma }_{t}$ in (\ref{eq:Sigma_t}) and $n_{t+1}$ in (\ref{eq:n_t+1}) until $t=PM$, the switch matrix $\bf S$ can be designed in a column-by-column way, which determines all locations of $M$ antennas in $P$ pilot timeslots. To avoid the high computational complexity of matrix inversion in (\ref{eq:Sigma_t}), the following lemma is derived to efficiently calculate ${\bf \Sigma}_t$ in a recursive way:
\begin{lemma}[Recursion formula of ${\bf \Sigma}_t$]
Given ${\bf \Sigma}_{t-1}$ and $n_t$, the posterior covariance ${\bf \Sigma}_{t}$ can be calculated by 
\begin{align}\label{eq:Sigma_t_recursion}
{{\bf{\Sigma }}_t} = \;\;{{\bf{\Sigma }}_{t - 1}} - {{{{\bf{\Sigma }}_{t - 1}}\left( {:,{n_t}} \right){{\bf{\Sigma }}_{t - 1}}\left( {{n_t},:} \right)} \over {{{\bf{\Sigma }}_{t - 1}}\left( {{n_t},{n_t}} \right) + {\sigma ^2}}}.
\end{align}
\end{lemma}
\begin{IEEEproof}
Constructive proof is given in Appendix B.	
\end{IEEEproof}
Then, for given $\bf S$, the optimal weight for reconstructing $\bf h$ can be obtained by 
\begin{equation}\label{eq:weight_W}
{\bf W}={\left( {{{\bf{S}}^{\rm{H}}}{\bf{\Sigma S}} + {\sigma ^2}{{\bf{I}}_{PM}}} \right)^{ - 1}}{{\bf{S}}^{\rm{H}}}{\bf{\Sigma }}.
\end{equation}

\subsubsection{Stage 2 (Online Regression)}
Since {\it Stage 1} is realized offline, the switch matrix ${\bf S}$ and the weight ${\bf W}$ can be designed and saved at the \ac{bs} in advance. This mechanism of offline design and online deployment can significantly reduce the computational complexity of employing S-BAR. In {\it Stage 2}, the scheme is then employed online for channel measurements. The $M$ fluid antennas of the \ac{bs} will move and receive pilots according to the designed ${\bf S}$, arriving at the noisy pilot $\bf y$. According to the Bayesian estimator in (\ref{eqn:mu_O}), channel $\bf h$ can be reconstructed by the weighted sum $\hat{\bf h} = {\bf W}^{\rm H}{\bf y}$, which finally completes the proposed S-BAR in {\bf Algorithm \ref{alg:SBAR}}.

\section{Performance Analysis and Kernel Selection}\label{sec:PA_KS}

\subsection{Performance Analysis of S-BAR}\label{subsec:PA}
To evaluate the performance limits of our proposed S-BAR, in this subsection, the estimation accuracy and computational complexity of S-BAR are analyzed, respectively. 

\subsubsection{Estimation Accuracy of S-BAR}
To quantitatively depict the estimation accuracy of S-BAR, we first define the \ac{mse} of reconstructing $\bf h$ as $E={\mathsf E}_{{\bf h},{\bf z}}( {{{\| {{\hat{\bf h}} - {\bf{h}}} \|}^2}} )$. By adopting some matrix techniques, we obtain the following lemma.
\begin{lemma}[\ac{mse} of S-BAR for given kernel $\bm \Sigma$]
Assume ${\mathsf E}_{\bf h}\left({\bf h}{\bf h}^{\rm H}\right) = {\bf{\Sigma }}_{{\rm cov}}$. Given a kernel $\bf \Sigma$ and a switch matrix $\bf S$ as the inputs of S-BAR, the \ac{mse} of reconstructing $\bf h$ can be derived as:
\begin{align}
\notag
E = &
{\rm{Tr}}\left( {{{\bf{\Pi }}^{\rm{H}}}\left( {{\bf{S}}^{\rm{H}}{{\bf{\Sigma }}_{{\mathop{\rm cov}} }}{{\bf{S}}} + {\sigma ^2}{{\bf{I}}_{PM}}} \right){\bf{\Pi }}} \right) - \\ \label{eqn:mse} &~~~~~~~~~~~~ 2\Re\left( {{\rm{Tr}}\left( {{{\bf{\Pi }}^{\rm{H}}}{\bf{S}}^{\rm{H}}{{\bf{\Sigma }}_{{\mathop{\rm cov}} }}} \right)} \right) + {\rm{Tr}}\left( {{{\bf{\Sigma }}_{{\mathop{\rm cov}} }}} \right),
\end{align}
where ${\bf{\Pi }}$ is a matrix function w.r.t $\bm \Sigma$, given by
\begin{align}\label{eqn:Pi}
{\bf{\Pi }} = {\left( {{\bf{S}}^{\rm H}{\bf \Sigma}{{\bf{S}}} + {\sigma ^2}{{\bf{I}}_{PM}}} \right)^{ - 1}}{\bf{S^{\rm H}\Sigma }}.
\end{align}
\end{lemma}
\begin{IEEEproof}
Constructive proof is given in Appendix C.
\end{IEEEproof}
{\bf Lemma 3} characterizes the estimation accuracy of S-BAR for a given kernel $\bm \Sigma$. Particularly, it allows us to evaluate the estimation error of S-BAR when the selected kernel $\bm \Sigma$ and the real covariance ${\bf{\Sigma }}_{{\rm cov}}$ are mismatched. Besides, we find that $E$ is independent of the real mean of channels, i.e., ${\mathsf E}_{\bf h}({\bf h})$, which means modeling $\bf h$ as a zero-mean process does not loss the generality. Then, as the fundamental limit, one may be concerned with the achievable minimum \ac{mse} of S-BAR, thus we introduce the following corollary.
\begin{lemma}[Achievable Minimum \ac{mse} of S-BAR]
Assume ${\mathsf E}\left({\bf h}{\bf h}^{\rm H}\right) = {\bf{\Sigma }}_{{\rm cov}}$, where ${\bf{\Sigma }}_{{\rm cov}}$ is rank-$K$. The achievable minimum \ac{mse} of reconstructing $\bf h$ via S-BAR, i.e. $E_{\min} = \mathop {\min }\limits_{{\bm \Sigma},{\bf S}}  ~ {\mathsf E}_{{\bf h},{\bf z}}( {{{\| {{\hat{\bf h}} - {\bf{h}}} \|}^2}} )$, only depends on the eigenvalues of ${\bf{\Sigma }}_{{\rm cov}}$ and the noise power $\sigma^2$, which can be  written as:
\begin{align}
E_{\min}  = \sum\limits_{k = 1}^K {\frac{{{\lambda _k}{\sigma ^2}}}{{{\lambda _k} + {\sigma ^2}}}}, \label{eqn:min-MSE}
\end{align}
where $\{\lambda_1,\cdots,\lambda_K\}$ are the $K$ positive eigenvalues of ${\bf{\Sigma }}_{{\rm cov}}$. The equality holds if and only if the kernel is exactly the real covariance of $\bf h$, i.e., ${\bf{\Sigma }} = {\bf{\Sigma }}_{{\rm cov}}$, and the \ac{fas} channels are completely observed, i.e., $PM=N$.
\end{lemma}
\begin{IEEEproof}
	Constructive proof is given in Appendix D.
\end{IEEEproof}

{\bf Lemma 4} characterizes the lower bound of \ac{mse} while employing S-BAR for \ac{fas} channel estimation. Note that, when the dimension of $\bf S$ is $N\times N$, the reconstructor degenerates into a well-informed estimator. Furthermore, if the kernel $\bf \Sigma$ is chosen to equal the real channel covariance ${\bf \Sigma}_{\rm cov}$, the \ac{mse} of the Bayesian reconstructor will coincide with the well-known posterior covariance formula in linear estimation theory \cite{kay1993fundamentals}. From the perspective of statistical signal processing, {\bf Lemma 4} reveals the equivalence between a perfect kernel-based Bayesian reconstructor and a LMMSE estimator.  

\begin{table}[t]
	\centering
	\small
	\caption{Computational Complexity of Different Schemes.}
	\label{table:1}
	\setstretch{1.25}
	\begin{tabular}{|c|c|c|c|c|c|c|}
		\hline  
		\bf Scheme&\bf Computational complexity \\ 
		\hline  
		FAS-OMP & ${\cal O}\left(LPM{N^2}\right)$ \\ \hline
		FAS-ML &${\cal O}\left({I_o}PML\left( {PM + N} \right)\right)$ \\ \hline
		S-BAR (Stage 1) & ${\cal O}\left({P}{M}\left( {{P^2}{M^2} + NPM + {N^2}} \right)\right)$ \\ \hline
		S-BAR (Stage 2) & ${\cal O}\left(PMN\right)$\\	
		\hline  
	\end{tabular}
	\vspace*{-0em}
\end{table}

\subsubsection{Computational Complexity of S-BAR}
The entire procedures of the existing \ac{fas}-\ac{omp} reconstructor in {\bf Algorithm \ref{alg:OMP}} \cite{ma2023compressed} and \ac{fas}-\ac{ml} reconstructor in {\bf Algorithm \ref{alg:ML}} should be employed online. Different from these schemes, the proposed S-BAR incorporates a hybrid offline and online implementation process. Specifically, the signal processing of  S-BAR scheme is composed of two stages, the offline design at {\it Stage 1} and the online regression at {\it Stage 2}. At {\it Stage 1}, the computational complexity is dominated by the calculation of posterior covariance ${\bm \Sigma }_{t}$ and weight $\bf W$. According to (\ref{eq:Sigma_t_recursion}) and (\ref{eq:weight_W}), the complexity of {\it Stage 1} is ${\cal O}\left({P^2}{M^2}\left( {{P^2}{M^2} + NPM + {N^2}} \right)\right)$.  At {\it Stage 2}, the computational complexity is from the weighted sum of received pilot $\bf y$, i.e., $\hat{\bf h} = {\bf W}^{\rm H}{\bf y}$. Thus, the computational complexity of {\it Stage 2} is ${\cal O}\left(PMN\right)$. For comparison, here we summarize their computational complexities in Table I, wherein $I_o$ denotes the number of iterations required by \ac{fas}-\ac{ml} reconstructor.

Note that, although the complexity of {\it Stage 1} is high, {\it Stage 1} can be implemented offline in advance. Then, the calculated switch matrix $\bf S$ and weight $\bf W$ can be saved at the \ac{bs} for the subsequent online regression in {\it Stage 2}. Therefore, in practical applications, the effective complexity of employing S-BAR is only linear to the number of ports $N$. Besides, we point out that the designs of $\bf S$ and $\bf W$ do not depend on the specific user, which suggests that the proposed S-BAR scheme can be extended to multi-user case without difficulty. In addition, $\hat{\bf h} = {\bf W}^{\rm H}{\bf y}$ in {\it Stage 2} can be calculated in a parallel way, which means that the time complexity of S-BAR is dimension-independent. These encouraging features further enhance the expansibility and the practicality of our proposed S-BAR.

\subsection{Kernel Selection of S-BAR}\label{subsec:Kernel}
Selecting a proper kernel $\bm \Sigma$ for S-BAR is essential in building an effective regression model. $\bm \Sigma$ determines the shape and flexibility of the proposed S-BAR, which in turn affects its ability to capture patterns and make accurate reconstruction. Considering the localized correlation property of \ac{fas} channels, an ideal kernel should assign higher similarity to nearby ports and decrease influence rapidly with distance. Since the channel covariance does not change so frequently as channels, a well-designed kernel ${\bm \Sigma}$ can work for a long time. 

\subsubsection{Covariance Kernel}
Since the mathematical significance of kernel is the prior covariance, an ideal approach is to use the real covariance of $\bf h$ as the kernel for reconstruction, i.e., ${\bm \Sigma}_{\rm cov}={\mathsf E}\left({\bf h}{\bf h}^{\rm H}\right)$. Before employing S-BAR, we can train an approximated ${\bm \Sigma}_{\rm cov}$ based on the \ac{csi} knowledge in some existing channel datasets, which can be obtained by
\begin{equation}\label{eqn:cov_ker}
	{\bm \Sigma}_{\rm cov} \approx \frac{1}{R}\sum\limits_{r = 1}^R {{{\bf{h}}_r}{\bf{h}}_r^{\rm H}},
\end{equation}
where ${\bf{h}}_r$ is the $r$-th channel for training kernels and $R$ is the number of training timeslots. 

\subsubsection{Experiential Kernel}
In some scenarios where the explicit \ac{csi} is difficult to acquire, training an experiential kernel based on received pilots is more practical. Let ${\bf x}_n$ denote the position of the $n$-th port. After balancing the complexity and practicality, two experiential kernels are recommended:

\begin{itemize}
	\item {\bf Exponential Kernel}: The exponential kernel ${\bm \Sigma}_{\rm exp} $, also known as the Laplacian kernel, is the most popular choice in Bayesian regression, which is given by
	\begin{equation}
		{\bm \Sigma}_{\rm exp}(n,n') =  \exp(-\eta{\|{\bf x}_n-{\bf x}_{n'}\|^2})
	\end{equation}
	for all $n,n'\in\{1,\cdots,N\}$, where $\eta>0$ is an adjustable hyperparameter. The exponential kernel is not sensitive to outliers, thus it is suitable to reconstruct channels without obvious regularity.
	\item {\bf Bessel Kernel}:
	The Bessel kernel ${\bm \Sigma}_{\rm bes}$ is well-suited to model complex-valued data with oscillatory or periodic patterns. The kernel is given by
	\begin{equation}
		{\bm \Sigma}_{\rm bes}(n,n') = J_0\left(\eta{\|{\bf x}_n-{\bf x}_{n'}\|}\right)
	\end{equation}
	for all $n,n'\in\{1,\cdots,N\}$, $J_0$ is the zero-order Bessel function of the first kind. ${\bm \Sigma}_{\rm bes}$ has the flexibility to adapt to data that exhibits repeating fluctuations, thus it is suitable to reconstruct the channels with periodic patterns.
\end{itemize}

The hyperparameter $\eta$ is the key factor that influences the performance of experiential kernels. We propose to determine the value of $\eta$ via an \ac{ml}-based method, so that the selected experiential kernels can well approximate the real channel covariance. Specifically, we assume that $R$ channel realizations are utilized to train an experiential kernel $\bf \Sigma$ before employing S-BAR. By viewing ${\bf \Sigma}$ as a function of hyperparameter $\eta$, the \ac{ml} estimator of $\eta$ can be written as
\begin{align}\label{eq:eta_opt}
\{{\bf \Sigma}^\star,\eta^\star\} = \mathop{\arg\!\max}_{{\bf \Sigma}\in\{{\bm \Sigma}_{\rm exp},{\bm \Sigma}_{\rm bes}\},\,\eta>0}~~{\sum\limits_{r = 1}^R}\ln\left({\mathsf P}\left({\bf y}_r|\eta\right)\right), 
\end{align}
wherein the likelihood function is given by
\begin{align}
{\mathsf P}\left({\bf y}_r|\eta\right) = {{\exp \left( { - {\bf{y}}_r^{\rm{H}}{{\left( {{\bf{S}}_r^{\rm{H}}{\bf{\Sigma }}{{\bf{S}}_r} + {\sigma ^2}{{\bf{I}}_{PM}}} \right)}^{ - 1}}{{\bf{y}}_r}} \right)} \over {{\pi ^{PM}}\det \left( {{\bf{S}}_r^{\rm{H}}{\bf{\Sigma }}{{\bf{S}}_r} + {\sigma ^2}{{\bf{I}}_{PM}}} \right)}}
\end{align}
and $\bf \Sigma$ can be selected as ${\bm \Sigma}_{\rm exp}$ or ${\bm \Sigma}_{\rm bes}$; ${\bf y}_r:={{\bf{S}}_r}{{\bf{h}}_r} + {{\bf{z}}_r}\in{\mathbb C}^{PM}$ denotes the received pilot associated with the $r$-th training channel ${{\bf{h}}_r}$; and ${{\bf{S}}_r}$ is randomly selected from $\cal S$. Since $\eta$ is a positive scalar and $\bf \Sigma$ only has two candidates, a one-dimensional search can be adopted to obtain the optimal solution to problem (\ref{eq:eta_opt}).

\section{Simulation Results}\label{sec:sim}
In this section, simulation results are carried out to verify the effectiveness of the proposed S-BAR. Firstly, the simulation setup and benchmark schemes are specified in Subsection \ref{subsec:sim_setup}. Then, the estimation behavior of S-BAR is analyzed in Subsection \ref{subsec:behavior}. Next, The impact of noise on the estimation accuracy is shown in Subsection \ref{subsec:SNR}, and the influence of pilot overhead on the estimation accuracy is studied in Subsection \ref{subsec:P}. Finally, the impact of electromagnetic (EM) coupling on estimation accuracy is investigated in Subsection \ref{subsec:EM_C}.

\begin{table}[t]
	\centering
	\small
	\caption{Simulation Parameters of \ac{fas} Channels}
	\label{table:2}
	\setstretch{1.25}
	\begin{tabular}{|c|c|c|c|c|c|c|}
		\hline  
		\bf Parameter&\bf QuaDRiGa \cite{Stephan'14} & \bf SSC in (\ref{eqn:h})\\
		\hline  
		Carrier frequency $f_c$ & 3.5 GHz & 3.5 GHz \\ \hline
		Number of clusters & 23 & 9 \\ \hline
		Number of rays  & 20 & 100 \\ \hline
		Path gains & \cite[Table 7.7.1-2]{cdl} & ${\cal CN}(0,1)$ \\ \hline
		Incident angles & \cite[Table 7.7.1-2]{cdl} & ${\cal U}\left(-\pi,+\pi\right)$ \\ \hline
		Max. Angle spread & 5$^\circ$ & 5$^\circ$ \\ \hline
		Path delays & \cite[Table 7.7.1-2]{cdl} & $\backslash$ \\ \hline
		Max. Doppler shift & 10 Hz & $\backslash$ \\
		\hline  
	\end{tabular}
	\vspace*{-0em}
\end{table}

\subsection{Simulation Setup and Benchmarks}\label{subsec:sim_setup}
Since we have assumed the normalized transmit power, the receiver \ac{snr} is defined as ${\rm SNR}=\frac{{\mathsf E}\left(\|{\bf h}\|^2\right)}{\sigma^2}$, of which the default value is set to $20$ dB. Let $\hat{\bf h}$ denote the estimated value of channel $\bf h$. The performance is evaluated by the \ac{nmse}, which is defined as ${\rm NMSE}={\mathsf E}_{{\bf h},{\bf z}}\left(\frac{\|{\bf h}-\hat{\bf h}\|^2}{\|{\bf h}\|^2}\right)$.

\subsubsection{Simulation Setup}
Otherwise particularly specified, the number of \ac{fas} ports is set to $N=256$ and that of fluid antennas is set to $M=4$. The length of the fluid antenna array is set to $W=10\lambda$, thus the port spacing should be $d=\frac{W}{N-1}$. The number of pilots is set to $P=10$. To account for both the model-mismatched case and model-matched case for the existing parametric estimators, the simulations are provided based on both the QuaDRiGa channel model in \cite{Stephan'14} and the spatially-sparse clustered (SSC) channel model in (\ref{eqn:h}). The channel parameters in 3GPP
TR 38.901 \cite{cdl} is used to generate the QuaDRiGa channels, and the main values are set as shown in Table \ref{table:2}. For the kernel settings, we set the number of training timeslots as $R=100$ to train kernels. The hyperparameter $\eta$ is obtained via (\ref{eq:eta_opt}) to generate the exponential kernel ${\bm \Sigma}_{\rm exp}$ and the Bessel kernel ${\bm \Sigma}_{\rm bes}$.


\begin{figure*}[!t]
	\centering
	\hspace{-5mm}
	\subfigure{\includegraphics[width=0.97\textwidth]{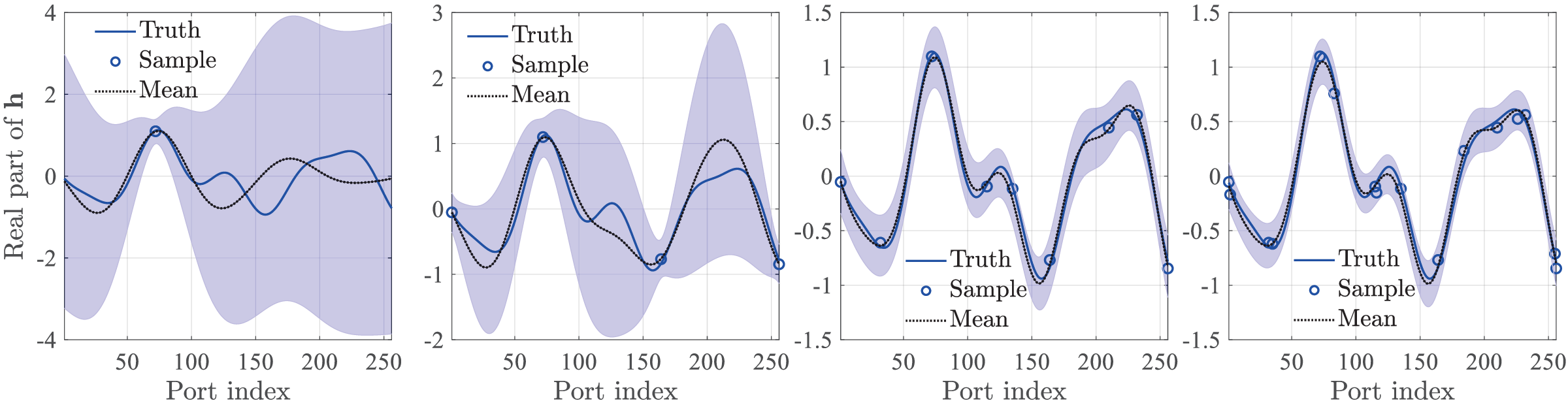}}
	\\
	\vspace{-0mm}
	\begin{flushleft}
		{\footnotesize \quad\quad\quad\quad\quad\quad(a) $P=1$, $M=1$.\quad\quad\quad\quad\quad\quad\quad(b) $P=2$, $M=2$.\quad\quad\quad\quad\quad\quad\quad(c) $P=3$, $M=3$.\quad\quad\quad\quad\quad\quad\quad(d) $P=4$, $M=4$.}
	\end{flushleft}
	\vspace{2mm}
	\hspace{-3mm}	
	\subfigure{\includegraphics[width=0.97\textwidth]{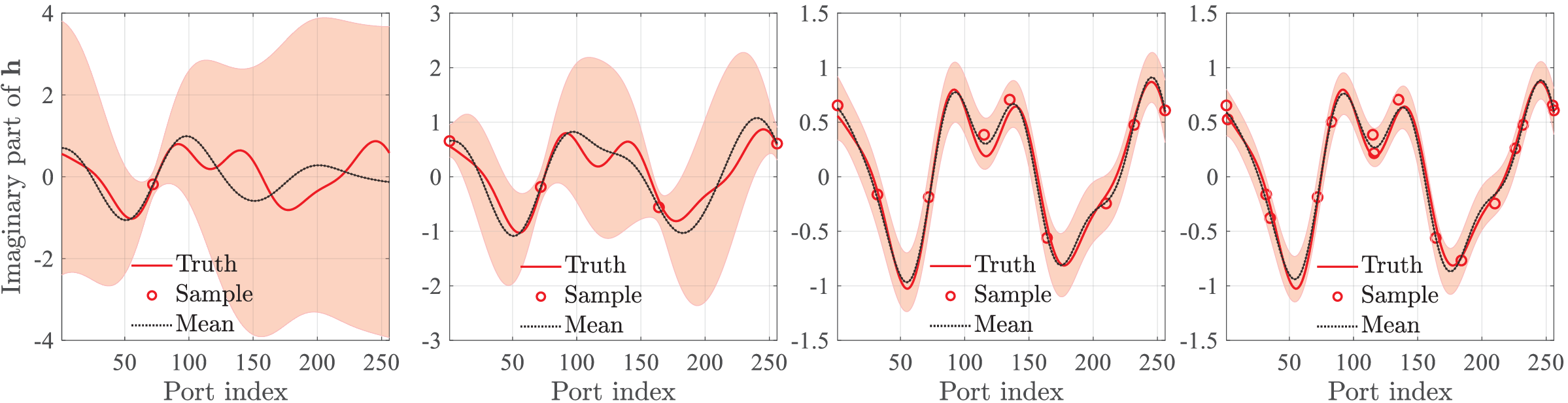}}
	\vspace{-0mm}
	\begin{flushleft}
		{\footnotesize \quad\quad\quad\quad\quad\quad(e) $P=1$, $M=1$.\quad\quad\quad\quad\quad\quad\quad(f) $P=2$, $M=2$.\quad\quad\quad\quad\quad\quad\quad(g) $P=3$, $M=3$.\quad\quad\quad\quad\quad\quad\quad(h) $P=4$, $M=4$.}
	\end{flushleft}
	\vspace*{-0.5em}
	\caption{ An illustration of employing S-BAR scheme to estimate \ac{fas} channel $\bf h$. (a)-(d) illustrate the real part of $\bf h$ versus the index of ports. (e)-(h) illustrate the imaginary part of $\bf h$ versus the index of ports.
	}
	\label{img:GPR}
\end{figure*}

\subsubsection{Simulation Schemes}

We consider the following four \ac{fas} channel estimators for simulations:
\begin{itemize}
	\item {\bf FAS-OMP:} Set the spatial sparsity $L$ as the twice of the number of clusters $C$. Then, {\bf Algorithm \ref{alg:OMP}}, i.e., the modified FAS-OMP reconstructor in \cite{ma2023compressed}, is employed to reconstruct $\bf h$.
	\item {\bf FAS-ML:}  Set the spatial sparsity as $L=2C$. Then, the FAS-ML reconstructor in {\bf Algorithm \ref{alg:ML}} is employed to estimate channel $\bf h$. In particular, the path gains and incident angles of $\bf h$ are initialized via \ac{omp}.
	\item {\bf SeLMMSE: } The SeLMMSE scheme proposed in \cite{Skouroumounis'23} is adopted to estimate channel $\bf h$. It can be achieved by measuring channels of $PM$ equally-spaced ports and then using zero-order interpolation to reconstruct $\bf h$.
	\item {\bf Proposed S-BAR ($\bm \Sigma$):} Given a kernel $\bm \Sigma$, {\bf Algorithm \ref{alg:SBAR}}, i.e., the proposed S-BAR, is employed to estimate $\bf h$.
\end{itemize}

\subsection{Estimation Behavior of the Proposed S-BAR}\label{subsec:behavior}

To better understand the working principle of the proposed S-BAR for channel estimation, we plot Fig. \ref{img:GPR} to intuitively show its behavior for different system parameters. The QuaDRiGa channel model is considered to generate channel $\bf h$. The ideal covariance kernel ${\bm \Sigma}_{\rm cov}$ is used as the input of S-BAR. Fig. \ref{img:GPR} (a)-(d) show the real part of $\bf h$ as a function of the port index, and Fig. \ref{img:GPR} (e)-(g) show the imaginary part of $\bf h$ as a function of the port index. Particularly, the curve ``Truth'' denotes the real channel $\bf h$, and the circle marks denote the sampled (measured) channels, which are selected by the switch matrix $\bf S$ designed via S-BAR. The black dotted line ``Mean'' denotes the posterior mean of Bayesian regression ${\bm \mu }_{{\bf h}|{\bf y}}$, i.e., the estimated channel $\hat{\bf h}$.  The highlighted shadows in the figures represent the confidence interval of $\bf h$, defined as $[ {\bm \mu }_{{\bf h}|{\bf y}}(n) -3{{\bm \Sigma }_{{\bf h}|{\bf y}}(n,n)},{\bm \mu }_{{\bf h}|{\bf y}}(n)+3{{\bm \Sigma }_{{\bf h}|{\bf y}}(n,n)}]$ for the $n$-th port channel.

From this figure, we have the following observations. Firstly, as $P$ and $M$ increase, the confidence interval, i.e., the vertical height of shadows, is gradually reduced. It indicates that, more pilots or antennas allow more sampling points for channel reconstruction, which can better eliminate the uncertainty of \ac{fas} channels. When the posterior variance ${{\bm \Sigma }_{{\bf h}|{\bf y}}(n,n)}$ becomes sufficiently small, the posterior mean $\mu_{\Omega}$ can well approximate $\bf h$. Secondly, Fig. \ref{img:GPR} (c) and Fig. \ref{img:GPR} (g) show that, nine samples are enough for the posterior mean ${\bm \mu }_{\Omega}$ to well approximate a 256-dimensional $\bf h$. This result suggests that, benefiting from the prior knowledge of strongly correlated channels, very few pilots and fluid antennas are sufficient to accurately reconstruct channels.
Thirdly, one can note that the samples in Fig. \ref{img:GPR} (d) and Fig. \ref{img:GPR} (h) are close. The reason is that, the inflection regions of a function are more informative than its flat regions, which indicates that the inflection parts require more dense samples to realize well regression. That's why the shadow heights in inflection regions are larger than those in flat regions, as shown in Fig. \ref{img:GPR} (c) and Fig. \ref{img:GPR} (g). Although some ports in inflection regions have been selected as samples, they are still hard to sufficiently eliminate the channel uncertainty of these regions. Thereby, in Fig. \ref{img:GPR} (d) and Fig. \ref{img:GPR} (h), some samples in inflection regions are selected to further eliminate the channel uncertainty of these regions.



\subsection{NMSE versus the Receiver SNR}\label{subsec:SNR}
\begin{figure}[!t]
	\centering
	\includegraphics[width=0.51\textwidth]{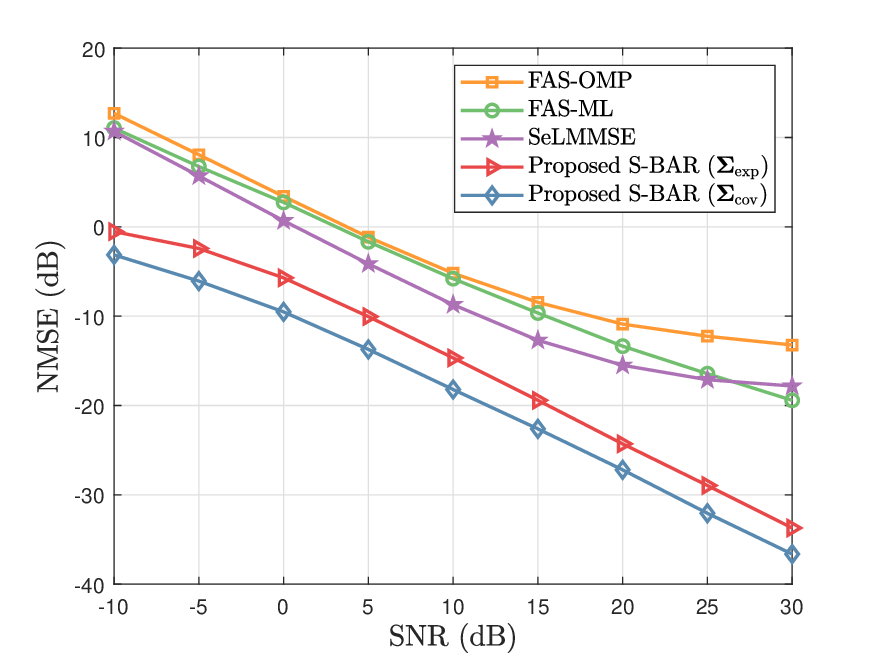}
	\caption{The \ac{nmse} as a function of the receiver \ac{snr} under the assumption of QuaDRiGa channel model.}
	\label{img:NMSE_vs_SNR}
\end{figure}
\begin{figure}[!t]
	\centering
	\includegraphics[width=0.51\textwidth]{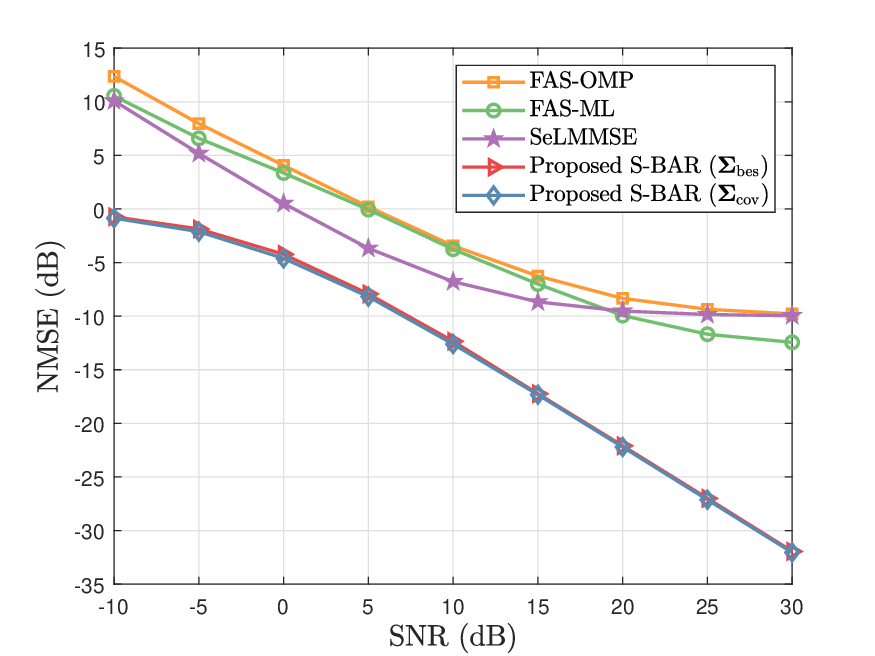}
	\caption{The \ac{nmse} as a function of the receiver \ac{snr} under the assumption of SSC channel model.}
	\label{img:NMSE_vs_SNR_SV}
\end{figure}

We plot the \ac{nmse} as a function of the \ac{snr} in Fig. \ref{img:NMSE_vs_SNR} and Fig. \ref{img:NMSE_vs_SNR_SV}, which follow the assumptions of QuaDRiGa model and SSC model in (\ref{eqn:h}), respectively.
Note that, the \ac{fas}-\ac{omp} and \ac{fas}-\ac{ml} reconstructors have assumed the SSC model in (\ref{eqn:h}). Due to the lack of obvious regularity for QuaDRiGa channels, the exponential kernel ${\bm \Sigma}_{\rm exp}$ is selected as the input of S-BAR in the QuaDRiGa case. Due to the periodic patterns of SSC channels in the spatial domain, the Bessel kernel ${\bm \Sigma}_{\rm bes}$ is selected as the input of S-BAR in the SSC case \cite{wong2021fluid,wong2020fluid}. To show the performance of S-BAR in the ideal case, the pre-trained covariance kernel ${\bm \Sigma}_{\rm cov}$ is considered in both cases. From these two figures, we have the following observations.

Firstly, as the receiver \ac{snr} rises, the \acp{nmse} of all estimators decrease rapidly. Particularly, one can find that the proposed S-BAR achieves the highest estimation accuracy for both SSC channels and QuaDRiGa channels. The reason is that, the existing methods do not utilize the prior knowledge of \ac{fas} channels for estimation. For FAS-OMP and FAS-ML reconstructors, the measured ports are randomly selected, which means that the information provided by the measured channels may not capture all patterns of $\bf h$. For SeLMMSE, although channel correlation is partially exploited, the unmeasured channels are obtained by zero-order interpolation. From the statistical perspective, the potential estimation errors of the unmeasured channels are not considered by SeLMMSE. In contrast, the proposed S-BAR has incorporated the effect of prior information into its estimator, which can reduce the potential estimation errors of all channels simultaneously. Besides, S-BAR does not assume a specific channel model, which implies that it is not influenced by the model mismatch as the FAS-OMP and FAS-ML reconstructors.

Secondly, we observe that the S-BAR enabled by the exponential kernel ${\bm \Sigma}_{\rm exp}$ and Bessel kernel ${\bm \Sigma}_{\rm bes}$ can achieve similar performance as the covariance kernel ${\bm \Sigma}_{\rm cov}$. In particular, the curves of the two S-BAR schemes are almost coincident in the SSC channel case. Recall that ${\bm \Sigma}_{\rm exp}$ and ${\bm \Sigma}_{\rm bes}$ are generated by experiential parameters, while ${\bm \Sigma}_{\rm cov}$ is trained from real channel data. This observation indicates that, even if the real channel covariance ${\bm \Sigma}_{\rm cov}$ is totally unknown, the experiential kernel ${\bm \Sigma}_{\rm exp}$ and ${\bm \Sigma}_{\rm bes}$ can still enable S-BAR to achieve considerable performance. In other words, to estimate channels accurately, the proposed S-BAR only needs the ``virtual'' prior knowledge provided by carefully selected experiential kernels, while the ``real'' channel covariance is actually not necessary. 

\subsection{NMSE versus the Number of Pilots}\label{subsec:P}

\begin{figure}[!t]
	\centering
	\includegraphics[width=0.51\textwidth]{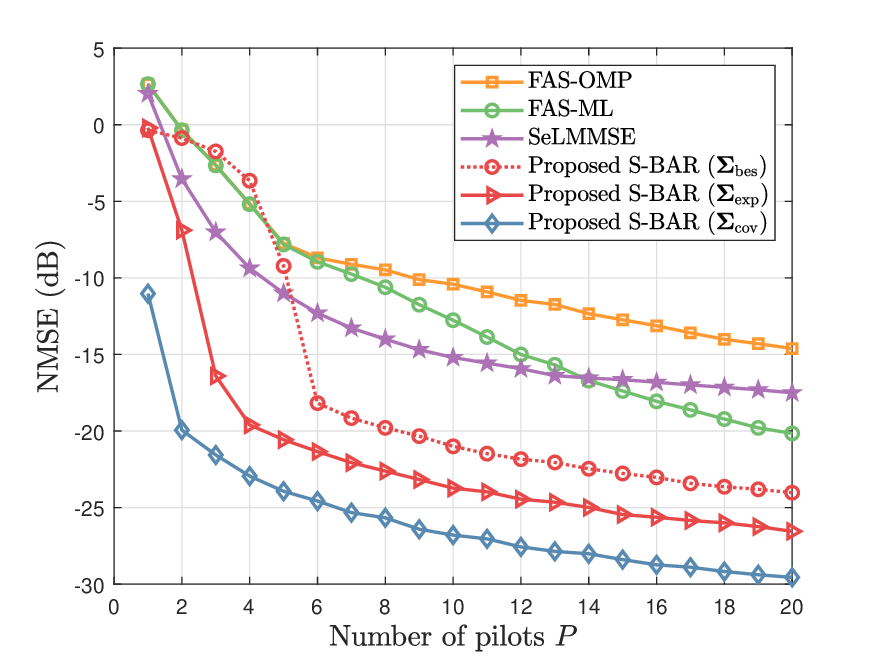}
	\caption{The \ac{nmse} as a function of the number of pilots $P$ under the assumption of QuaDRiGa channel model.}
	\label{img:NMSE_vs_P}
\end{figure}
\begin{figure}[!t]
	\centering
	\includegraphics[width=0.51\textwidth]{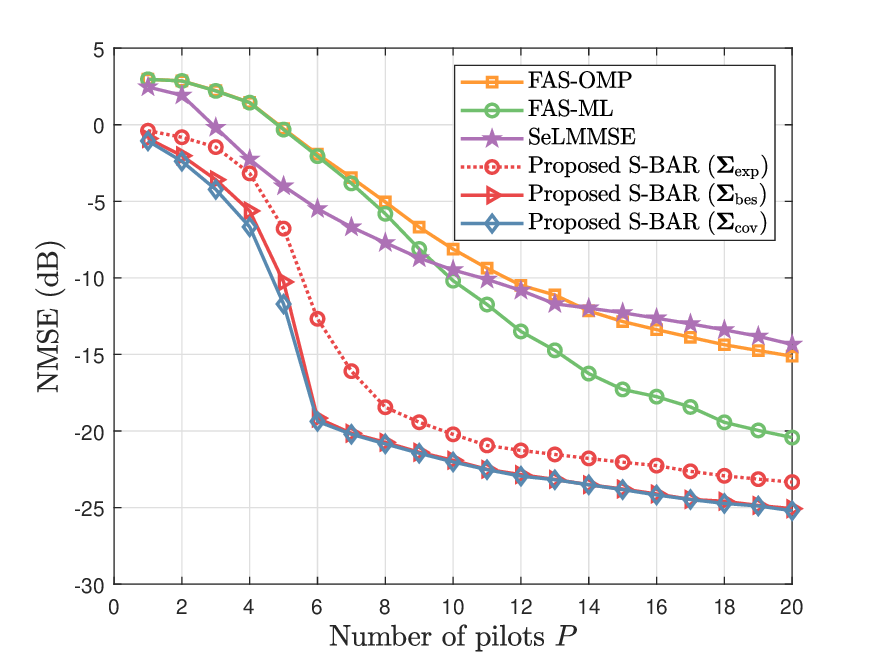}
	\caption{The \ac{nmse} as a function of the number of pilots $P$ under the assumption of SSC channel model.}
	\label{img:NMSE_vs_P_SV}
\end{figure}

We plot the \ac{nmse} as a function of the number of pilots $P$ in Fig. \ref{img:NMSE_vs_P} and Fig. \ref{img:NMSE_vs_P_SV}, which follow the assumptions of QuaDRiGa model and SSC model, respectively. In addition, to show the influence of non-preferred kernels on S-BAR in the two channel cases, we have also added the baseline that ${\bf \Sigma}_{\rm bes}$ is selected as the input of S-BAR in Fig. \ref{img:NMSE_vs_P} and the baseline that ${\bf \Sigma}_{\rm exp}$ is selected as the input of S-BAR in Fig. \ref{img:NMSE_vs_P_SV}. These new baselines are expressed as dotted red lines in the figures.

We observe from these two figures that, to achieve the same estimation accuracy, the proposed S-BAR scheme consumes a much lower pilot overhead than the benchmark schemes. For example, to achieve an \ac{nmse} of -15 dB in the SSC channel case, the numbers of pilots required by \ac{fas}-\ac{omp},  \ac{fas}-\ac{ml}, SeLMMSE, and the proposed S-BAR with the preferred kernel input are $P = 20$, $13$, $20$, and $5$, respectively. We can conclude that, compared to the state-of-art schemes, the proposed S-BAR scheme can reduce the pilot overhead of about 50\%. Particularly, in Fig. \ref{img:NMSE_vs_P}, one can find that the curve ``Proposed S-BAR (${\bf \Sigma}_{\rm cov}$)'' begins from ${\rm NMSE}=-10$ dB when $P=1$. It implies that, aided by kernel ${\bf \Sigma}_{\rm cov}$, the acquired information of four observations ($P\times M = 4$) is already sufficient for S-BAR to accurately recover the channels of most ports.


Besides, we also find that, even if the kernel input is not preferred, the proposed S-BAR can still achieve satisfactory performance. For example, when ${\bf \Sigma}_{\rm bes}$ is selected as the input of S-BAR in the QuaDRiGa channel case, the \ac{nmse} for S-BAR can achieve -21 dB when $P=10$, which is 6 dB lower than that for the best-performed SeLMMSE. In the SSC channel case, when ${\bf \Sigma}_{\rm exp}$ is selected as the input of S-BAR, its \ac{nmse} can achieve -20 dB, which is about 10 dB lower than the \ac{nmse} for the best-performed FAS-ML. These results have further verified the effectiveness of our proposed S-BAR.

Finally, in the SSC channel case, the \ac{nmse} for S-BAR decreases quickly when $P<6$, while it decreases slowly when $P>6$. The reason is that, when $P<6$, the proposed S-BAR has not captured all spatial features of SSC channels, thus the estimation error is mainly caused by the incomplete regression. When $P>6$, with sufficient channel sampling, the channel information captured by S-BAR can well describe the periodic feature of SSC channels. In this case, the estimation error is mainly caused by the measurement noise. On the other hand, in the QuaDRiGa channel case, this interesting phenomenon is not observed. The reason is that, unlike the SSC channels, the spatial periodicity is not obvious for the QuaDRiGa channels, which makes it difficult for S-BAR to fully capture all patterns of channels via a few samples. It explains why ${\bf \Sigma}_{\rm bes}$ is not the preferred kernel for the QuaDRiGa channel case.

\subsection{Impact of EM Coupling on Estimation Accuracy}\label{subsec:EM_C}
\begin{figure}[!t]
	\centering
	\includegraphics[width=0.51\textwidth]{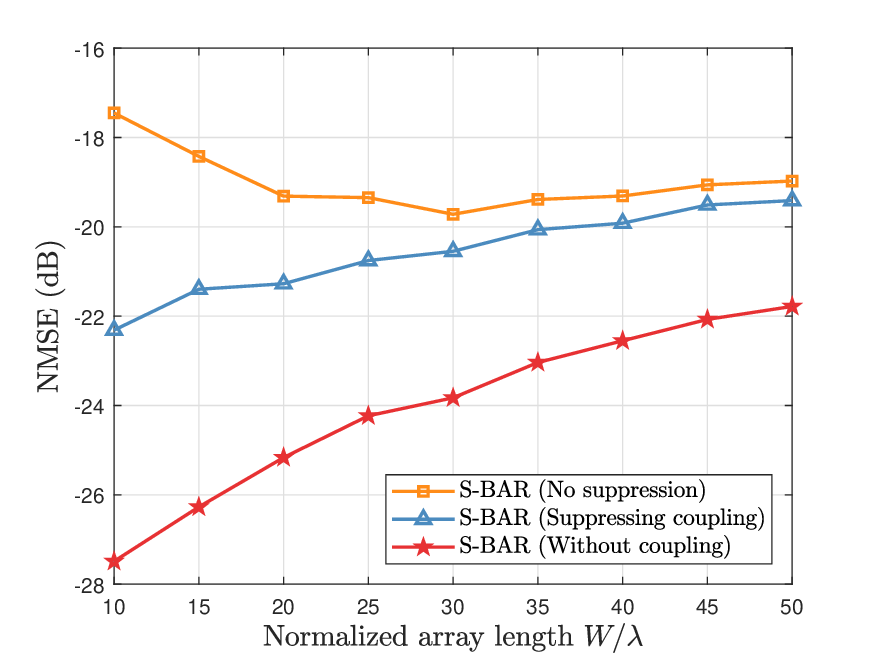}
	\caption{The \ac{nmse} as a function of the normalized array length $W/\lambda$.}
	\label{img:NMSE_vs_W}
\end{figure}
For some FASs whose multiple antennas can move physically, the effect of EM coupling among fluid antennas can be avoided by forcing that the antenna spacing is always larger than $\lambda/2$ \cite{wu2023movable,xiao2023multiuser}. In contrast, a few FASs are realized by pixel-like switch networks \cite{Daniel'14}. For such FAS structures, the EM coupling among ports cannot be ignored, thus only the coupled channels can be estimated. To take this effect into account, we introduce the EM coupling model in \cite{Svantesson'01}.  To be specific, coupling matrix ${\bf C}_{\rm em}\in{\mathbb C}^{N\times N}$ is used to modify the FAS channel $\bf h$, which can be obtained by field tests or full-wave simulations. As a typical example, here we consider 
dipole antennas with $\lambda/2$ length, $\lambda/100$ width, and $d:=\frac{W}{N-1}$ spacing. With these parameters, the desired ${\bf C}_{\rm em}$ can be directly generated by the Matlab Antenna Toolbox. In this context, the equivalent channel can be modeled as $\bar{\bf h}={\bf C}_{\rm em}^{1/2}{\bf h}$, and the equivalent kernel can be derived as ${\bf C}_{\rm em}^{1/2}{\bf \Sigma}({\bf C}_{\rm em}^{1/2})^{\rm H}$. To show the effectiveness of S-BAR against the EM coupling effect, three schemes are compared:
\begin{itemize}
	\item {\bf No suppression:} Under the influence of EM coupling, the proposed S-BAR ignores coupling and estimates the equivalent channel $\bar{\bf h}$ based on the original kernel ${\bf \Sigma}_{\rm cov}$.
	\item {\bf Suppressing coupling:} Under the influence of EM coupling, the proposed S-BAR is employed to estimate $\bar{\bf h}$ base on the modified kernel ${\bf C}_{\rm em}^{1/2}{\bf \Sigma}_{\rm cov}({\bf C}_{\rm em}^{1/2})^{\rm H}$.
	\item {\bf Without coupling:}  
	Assume that the inter-port EM coupling can be fully eliminated by hardware isolation, i.e., ${\bf C}_{\rm em}={\bf I}_N$. The proposed S-BAR is employed to estimate ideal ${\bf h}$ based on the ideal kernel ${\bf \Sigma}_{\rm cov}$.
\end{itemize}

Since the EM coupling depends on the port spacing $d$, we fix $M$ and then plot the NMSE versus the normalized array length ${W}/\lambda$ in Fig. \ref{img:NMSE_vs_W}, where the QuaDRiGa channels are considered. We observe that, despite the EM coupling among ports, the proposed S-BAR can still achieve considerable estimation performances, and the NMSE is still lower than -15 dB. It is interesting to find that, as the port spacing increases,  the NMSE for the scheme ``No suppression'' decreases first and then rises. The reason is that, the estimation performance of S-BAR is influenced by the two factors, i.e., spatial correlation and EM coupling. When $W<30\lambda$, the main factor that limits the estimation performance of ``No suppression'' scheme is EM coupling. As the port spacing increases, the effect of EM coupling is reduced, leading to its higher estimation accuracy. When $W>30\lambda$, the main factor becomes the spatial correlation. As the port spacing increases, the spatial correlation among ports becomes weaker. Since the accuracy improvement of S-BAR is realized by utilizing the strong correlation among ports, its performance gains become less accordingly. In contrast, the effect of EM coupling has been eliminated by ``Suppressing coupling'' through considering the equivalent kernel ${\bf C}_{\rm em}^{1/2}{\bf \Sigma}_{\rm cov}({\bf C}_{\rm em}^{1/2})^{\rm H}$, and ${\bf C}_{\rm em}$ is not taken into account by ``Without coupling'' case. Thus, the accuracy of these two estimators mainly depends on the spatial correlation. As the port spacing increases, the weaker spatial correlation becomes the bottleneck limiting their performance, which leads to their continuously rising NMSEs.

\section{Conclusions}\label{sec:con}
In this paper, we have proposed S-BAR as a general solution to estimate channels in \acp{fas}.  Specifically, this paper is the first attempt to introduce Bayesian inference into the FAS channel estimation. Different from the existing \ac{fas} channel estimators relying on channel assumptions, the general S-BAR utilizes the experiential kernel to estimate channels in a non-parametric way. Inspired by the Bayesian regression, the proposed S-BAR can select a few informative channels for measurement and combine them with the experiential kernel to reconstruct high-dimensional \ac{fas} channels. Simulation results reveal that, in both model-mismatched and model-matched cases, the proposed S-BAR can achieve higher estimation accuracy than the existing schemes relying on channel assumptions. 

For the follow-up works, the extension to wideband channel estimation for \acp{fas} will be interesting. Since the channels of multiple carriers share the same ports, the port selection should balance the uncertainty of the channels on different subcarriers.
Besides, thanks to the generality of Bayesian regression, the proposed S-BAR can be extended to realize channel estimation in the general phased-array case \cite{Cui'24'TIT}. In addition, since reconfigurable intelligent surfaces (RISs) can provide additional control degrees of freedom (DoFs) for FAS channels \cite{Zhang'21}, the cooperation between RISs and FASs may be able to improve the accuracy of \ac{csi} acquisitions \cite{Shojaeifard'22}. 


\appendices

\section{Proof of {\bf Lemma 1}}
Using some matrix partition operations, the mutual information $I({\bf y}_{t+1};{\bf h})$ can be rewritten as
\begin{align}
	&I({\bf y}_{t+1};{\bf h}) = \log_2 \det\left({\bf I}_{t+1} + 
	\frac{1}{\sigma^2}
	{\bf S}_{t+1}^{\rm H}{\bm \Sigma} {\bf S}_{t+1} 
	\right)\notag \\
	& = \log_2\det \left(
	\begin{array}{cc}
		{\bf I}_t + \frac{1}{\sigma^2}{\bf S}_t^{\rm H}{\bm \Sigma}{\bf S}_t & 
		\frac{1}{\sigma^2}{\bf S}_t^{\rm H}{\bm \Sigma} {\bf s}_{t + 1} \\
		\frac{1}{\sigma^2}{\bf s}_{t + 1}^{\rm H} {\bm \Sigma}{\bf S}_t & 
		1 + \frac{1}{\sigma^2}{\bf s}_{t + 1}^H {\bm \Sigma}{\bf s}_{t + 
			1}  
	\end{array}
	\right)
	\notag \\
	&\overset{(a)}{=} \log_2\det\left(
	\begin{array}{cc}
		{\bf I}_t + \frac{1}{\sigma^2}{\bf S}_t^{\rm H}{\bm \Sigma}{\bf S}_t & 
		\frac{1}{\sigma^2}{\bf S}_t^{\rm H}{\bm \Sigma} {\bf s}_{t + 1} \\
		{\bf 0}_{1\times t} & 
		1 + \frac{1}{\sigma^2}{\bf s}_{t + 1}^{\rm H} {\bm \Sigma}_t{\bf s}_{t + 
			1}  
	\end{array}
	\right)
	\notag \\ 
	& \overset{(b)}{=} I({\bf y}_t; {\bf h}) + \log_2 \left(1 + \frac{1}{\sigma^2}{\bf s}_{t + 
		1}^{\rm H}
	{\bm \Sigma}_{t}{\bf s}_{t + 1} \right),
\end{align}
where $(a)$ holds according to the matrix triangularization:
\begin{align}
	\notag &
\left[
\begin{array}{cc}
	{\bf I}_t + \frac{1}{\sigma^2}{\bf S}_t^{\rm H}{\bm \Sigma}{\bf S}_t & 
	\frac{1}{\sigma^2}{\bf S}_t^{\rm H}{\bm \Sigma} {\bf s}_{t + 1} \\
	{\bf 0}_{1\times t} & 
	1 + \frac{1}{\sigma^2}{\bf s}_{t + 1}^{\rm H} {\bm \Sigma}_t{\bf s}_{t + 
		1}  
\end{array} 
\right]
= \\ \notag & ~~~~~~~~
\left[
\begin{array}{cc}
{{{\bf{I}}_t}} & {{{\bf{0}}_t}}  \\ 
{ - {\bf{s}}_{t + 1}^{\rm{H}}{\bf{\Sigma }}{{\bf{S}}_t}{{\left( {{\bf{S}}_t^{\rm{H}}{\bf{\Sigma }}{{\bf{S}}_t} + {\sigma ^2}{{\bf{I}}_t}} \right)}^{ - 1}}} & 1  \\ 
\end{array}
\right] \times \\ &
\left[
\begin{array}{cc}
	{\bf I}_t + \frac{1}{\sigma^2}{\bf S}_t^{\rm H}{\bm \Sigma}{\bf S}_t & 
	\frac{1}{\sigma^2}{\bf S}_t^{\rm H}{\bm \Sigma} {\bf s}_{t + 1} \\
	\frac{1}{\sigma^2}{\bf s}_{t + 1}^{\rm H} {\bm \Sigma}{\bf S}_t & 
	1 + \frac{1}{\sigma^2}{\bf s}_{t + 1}^H {\bm \Sigma}{\bf s}_{t + 
		1}  
\end{array}
\right],
\end{align}
and $(b)$ holds according to the definition in (\ref{eq:MI_y_t}). Clearly, maximizing  $I({\bf y}_{t+1}; {\bf h}) - I({{\bf y}_t; 
	{\bf h}}) = \log_2 \left(1 + \frac{1}{\sigma^2}{\bf s}_{t + 1}^{\rm H} 
{\bm \Sigma}_{t}{\bf s}_{t + 1} \right)$ is equivalent to finding a proper $\bf s$ that maximizes ${\bf s}^{\rm H} {\bm \Sigma}_{t}{\bf s}$, which completes the proof.

\begin{figure*}[!b]
	\normalsize	
	\hrulefill
	\begin{align} \label{eq:block-matrix-inv}
		\left[
		\begin{array}{cc}
			{\bf A}	&  	{\bf B}
			\\ 
			{\bf C}	&		{\bf D}
		\end{array}\right]^{-1} = 	\left[
		\begin{array}{cc}
			{\bf A}^{-1} + {\bf A}^{-1} {\bf B}({\bf D} - {\bf C }{\bf 
				A}^{-1} {\bf 
				B})^{-1}{\bf C}{\bf A}^{-1}	&  	-{\bf A}^{-1}{\bf B}({\bf 
				D} - {\bf C}{\bf 
				A}^{-1}{\bf B})^{-1}
			\\ 
			-({\bf D} - {\bf C}{\bf A}^{-1}{\bf B})^{-1}{\bf C} {\bf 
				A}^{-1}	
			&			({\bf D} - {\bf C}{\bf A}^{-1}{\bf 
				B})^{-1} 
		\end{array}\right]
	\end{align} 
\end{figure*}

\section{Proof of {\bf Lemma 2}}

To obtain the recursion formula of ${\bf \Sigma}_t$, the key step is to use block-matrix inversion formula in (\ref{eq:block-matrix-inv}) at the bottom of the next page to process the term ${\left( {{\bf{S}}_t^{\rm{H}}{\bf{\Sigma }}{{\bf{S}}_t} + {\sigma ^2}{{\bf{I}}_t}} \right)^{ - 1}}$ in (\ref{eq:Sigma_t}). Letting ${\bf \Xi}_t={{\bf{S}}_t^{\rm{H}}{\bf{\Sigma }}{{\bf{S}}_t} + {\sigma ^2}{{\bf{I}}_t}}$, we have 
\begin{align}\label{eq:ABCD}
{\bf \Xi}_t = 
\left[
\begin{array}{cc}
	{\bf \Xi}_{t-1} & 
	{\bf S}_{t-1}^{\rm H}{\bm \Sigma} {\bf s}_{t} \\
{\bf s}_{t}^{\rm H} {\bm \Sigma}{\bf S}_{t-1} & 
	\sigma^2 + {\bf s}_{t}^H {\bm \Sigma}{\bf s}_{t}  
\end{array}
\right]=\left[
\begin{array}{cc}
	{\bf A}	&  	{\bf B}
	\\ 
	{\bf C}	&		{\bf D}
\end{array}\right].
\end{align}
Substituting the corresponding $\bf A$, $\bf B$, $\bf C$, and $\bf D$ in (\ref{eq:ABCD}) into (\ref{eq:block-matrix-inv}), the inverse of ${\bf \Xi}_t$ can be represented by the combination of ${\bf \Xi}_{t-1}^{-1}$, ${\bf S}_{t-1}$, and ${\bf s}_{t}$, written as
\begin{align}
{\bf{\Xi }}_t^{ - 1} \!=\! \left[ {\begin{array}{*{20}{c}}
		{{\bf{\Xi }}_{t - 1}^{ - 1} \!+\! \frac{{{\bf{\Xi }}_{t - 1}^{ - 1}{\bf{S}}_{t - 1}^{\rm{H}}{\bf{\Sigma }}{{\bf{s}}_t}{\bf{s}}_t^{\rm{H}}{\bf{\Sigma }}{{\bf{S}}_{t - 1}}{\bf{\Xi }}_{t - 1}^{ - 1}}}{{{\sigma ^2} + {\bf{s}}_t^H{{\bf{\Sigma }}_{t - 1}}{{\bf{s}}_t}}}}&{\frac{{ - {\bf{\Xi }}_{t - 1}^{ - 1}{\bf{S}}_{t - 1}^{\rm{H}}{\bf{\Sigma }}{{\bf{s}}_t}}}{{{\sigma ^2} + {\bf{s}}_t^H{{\bf{\Sigma }}_{t - 1}}{{\bf{s}}_t}}}}\\
		{\frac{{ - {\bf{s}}_t^{\rm{H}}{\bf{\Sigma }}{{\bf{S}}_{t - 1}}{\bf{\Xi }}_{t - 1}^{ - 1}}}{{{\sigma ^2} + {\bf{s}}_t^H{{\bf{\Sigma }}_{t - 1}}{{\bf{s}}_t}}}}&{\frac{1}{{{\sigma ^2} + {\bf{s}}_t^H{{\bf{\Sigma }}_{t - 1}}{{\bf{s}}_t}}}}
\end{array}} \right].
\end{align}
Then, by taking ${\bf \Xi}_t^{-1}$ and ${\bf S}_t=[{\bf S}_{t-1}~{\bf s}_t]$ back into the posterior covariance ${\bf \Sigma}_t$ in (\ref{eq:Sigma_t}), we have 
\begin{align}
\notag
&{{\bf{\Sigma }}_t} = {\bf{\Sigma }} - {\bf{\Sigma }}{{\bf{S}}_t} {\bf \Xi}_t^{-1} {\bf{S}}_t^{\rm{H}}{\bf{\Sigma }} \\
& = {\bf{\Sigma }} - {\bf{\Sigma }}{{\bf{S}}_{t-1}} {\bf \Xi}_{t-1}^{-1} {\bf{S}}_{t-1}^{\rm{H}}{\bf{\Sigma }} + {\frac{1}{{{\sigma ^2} + {\bf{s}}_t^H{{\bf{\Sigma }}_{t - 1}}{{\bf{s}}_t}}}}\times \notag
\\ & {\Big (}  - \left( {{\bf{\Sigma }} - {{\bf{\Sigma }}_{t - 1}}} \right){{\bf{s}}_t}{\bf{s}}_t^{\rm{H}}\left( {{\bf{\Sigma }} - {{\bf{\Sigma }}_{t - 1}}} \right)  + \left( {{\bf{\Sigma }} - {{\bf{\Sigma }}_{t - 1}}} \right){{\bf{s}}_t}{\bf{s}}_t^{\rm{H}}{\bf{\Sigma }} \notag \\
& + {\bf{\Sigma }}{{\bf{s}}_t}{\bf{s}}_t^{\rm{H}}\left( {{\bf{\Sigma }} - {{\bf{\Sigma }}_{t - 1}}} \right) + {\bf{\Sigma }}{{\bf{s}}_t}{\bf{s}}_t^{\rm{H}}{\bf{\Sigma }}
{\Big )} \notag \\
& = {{\bf{\Sigma }}_{t - 1}} - \frac{{{{\bf{\Sigma }}_{t - 1}}{{\bf{s}}_t}{\bf{s}}_t^H{{\bf{\Sigma }}_{t - 1}}}}{{{\sigma ^2} + {\bf{s}}_t^H{{\bf{\Sigma }}_{t - 1}}{{\bf{s}}_t}}} \notag \\
& \overset{(a)}{=} {{\bf{\Sigma }}_{t - 1}} - {{{{\bf{\Sigma }}_{t - 1}}\left( {:,{n_t}} \right){{\bf{\Sigma }}_{t - 1}}\left( {{n_t},:} \right)} \over {{{\bf{\Sigma }}_{t - 1}}\left( {{n_t},{n_t}} \right) + {\sigma ^2}}},
\end{align}
where $(a)$ holds since ${\bf{s}}_t^H{{\bf{\Sigma }}_{t - 1}}{{\bf{s}}_t} = {{\bf{\Sigma }}_{t - 1}}\left( {{n_t},{n_t}} \right)$ and ${\bf{\Sigma }}{{\bf{s}}_t} = {{\bf{\Sigma }}_{t - 1}}\left( {:,{n_t}} \right)$. This completes the proof.

\section{Proof of {\bf Lemma 3}}
When $t = PM$, according to the definition of $\hat{\bf h}$ in (\ref{eq:channel_reconstrcutor}), the square error of employing S-BAR can be rewritten as
\begin{align}
{{{\left\| { \hat{\bf h} - {\bf{h}}} \right\|}^2}}
\notag
&\stackrel{(a)}{=}  {\Big \|} 
{\left( {{{\left( {{\bf{S^{\rm H}\Sigma }}} \right)}^{\rm{H}}}{{\left( {{\bf{S^{\rm H}\Sigma }}{{\bf{S}}} + {\sigma ^2}{{\bf{I}}_{PM}}} \right)}^{ - 1}}{\bf{S}}^{\rm H} - {{\bf{I}}_N}} \right){\bf{h}}} \\ \notag & ~~~~~~~~~~~+
{{{\left( {\bf{S}}^{\rm H}{\bf{\Sigma }} \right)}^{\rm{H}}}{{\left( {{\bf{S^{\rm H}\Sigma }}{{\bf{S}}} + {\sigma ^2}{{\bf{I}}_{PM}}} \right)}^{ - 1}}{\bf{z}}}
 {\Big \|}^2
 \\
& \stackrel{(b)}{=} {\left\| {\left( {{{\bf{\Pi }}^{\rm{H}}}{\bf{S}}^{\rm H} - {{\bf{I}}_N}} \right){\bf{h}} + {{\bf{\Pi }}^{\rm{H}}}{\bf{z}}} \right\|^2}
\end{align}
where $(a)$ holds according to (\ref{eqn:y}); $(b)$ holds according to (\ref{eqn:Pi}). Next, recalling the properties ${\bf z} \sim \mathcal{C} \mathcal{N}\!\left({\bf 0}_{PM}, \sigma^2{\bf I}_{PM} \right)$ and ${\mathsf E}\left({\bf h}{\bf h}^{\rm H}\right) = {\bf{\Sigma }}_{{\rm cov}}$, the \ac{mse} can be derived as
\begin{align}
\notag	
&E ={\mathsf E}\left( {{{\left\| {{{\bm{\mu }}_\Omega } - {\bf{h}}} \right\|}^2}} \right)	
\\
\notag 
& \stackrel{(c)}{=} {\rm{Tr}}\left( {\left( {{{\bf{\Pi }}^{\rm{H}}}{\bf{S}}^{\rm{H}} - {{\bf{I}}_N}} \right){{\bf{\Sigma }}_{{\mathop{\rm cov}} }}{{\left( {{{\bf{\Pi }}^{\rm{H}}}{\bf{S}}^{\rm{H}} - {{\bf{I}}_N}} \right)}^{\rm{H}}}} \right) + {\sigma ^2}{\rm{Tr}}\left( {{{\bf{\Pi }}^{\rm{H}}}{\bf{\Pi }}} \right)
\\
& \notag
={\rm{Tr}}\left( {{{\bf{\Pi }}^{\rm{H}}}\left( {{\bf{S}}^{\rm{H}}{{\bf{\Sigma }}_{{\mathop{\rm cov}} }}{{\bf{S}}} + {\sigma ^2}{{\bf{I}}_{PM}}} \right){\bf{\Pi }}} \right) - \\ &~~~~~~~~~~~~ 2\Re \left( {{\rm{Tr}}\left( {{{\bf{\Pi }}^{\rm{H}}}{\bf{S}}^{\rm{H}}{{\bf{\Sigma }}_{{\mathop{\rm cov}} }}} \right)} \right) + {\rm{Tr}}\left( {{{\bf{\Sigma }}_{{\mathop{\rm cov}} }}} \right),
\end{align}
where $(c)$ holds since ${\left\| {\bf{x}} \right\|^2} = {\rm{Tr}}\left( {{\bf{x}}{{\bf{x}}^{\rm H}}} \right)$ for any vector $\bf x$. This completes the proof.

\section{Proof of {\bf Lemma 4}}
Observing (\ref{eqn:mse}), one can find that the \ac{mse} $E$ is a quadratic form w.r.t matrix $\bm \Pi$. Thus, the minimum \ac{mse} can be achieved by finding an $\bm \Sigma$ such that $\frac{{\partial E}}{{\partial {\bf{\Pi }}}}={\bf 0}_{PM\times N}$ holds. According to (\ref{eqn:mse}), the partial derivative of $E$ with respect to $\bm \Pi$ can be derived as
\begin{align}\label{eqn:partial}
\frac{{\partial E}}{{\partial {\bf{\Pi }}}} = {\left( {{\bf{S}}^{\rm{H}}{{\bf{\Sigma }}_{{\mathop{\rm cov}} }}{{\bf{S}}} + {\sigma ^2}{{\bf{I}}_{PM}}} \right)^*}{{\bf{\Pi }}^*} - {\left( {{{\bf{\Sigma }}_{{\mathop{\rm cov}} }}{{\bf{S}}}} \right)^{\rm T}}.
\end{align}
By letting $\frac{{\partial E}}{{\partial {\bf{\Pi }}}}={\bf 0}_{PM\times N}$ and substituting (\ref{eqn:Pi}) into (\ref{eqn:partial}), the original proof can be reduced to prove that 
\begin{align}\label{eqn:proof}
\notag
&{\left( {{\bf{S^{\rm{H}}\Sigma }}{{\bf{S}}} + {\sigma ^2}{{\bf{I}}_{PM}}} \right)^{ - 1}}{\bf{S^{\rm{H}}\Sigma }} = \\ &~~~~~~~~ {\left( {{\bf{S}}^{\rm{H}}{{\bf{\Sigma }}_{{\mathop{\rm cov}} }}{{\bf{S}}} + {\sigma ^2}{{\bf{I}}_{PM}}} \right)^{ - 1}}{\bf{S}}^{\rm{H}}{{\bf{\Sigma }}_{{\mathop{\rm cov}} }}.
\end{align}
Since the proposed S-BAR does not depend on the value of noise power, (\ref{eqn:proof}) should hold for arbitrary $\sigma^2$. By considering the special case when $\sigma^2\to\infty$, (\ref{eqn:proof}) is equivalent to
\begin{align}\label{eqn:proof2}
{\bf{S}}^{\rm{H}}\left( {{\bf{\Sigma }} - {{\bf{\Sigma }}_{{\mathop{\rm cov}} }}} \right) = {{\bf{0}}_{PM \times N}}
\end{align}
Furthermore, since our derivation also does not impose the specific form of $\bf S$, ${{\bf{\Sigma }} = {{\bf{\Sigma }}_{{\mathop{\rm cov}} }}}$ is necessary to make the equation always hold. By substituting ${\bf{\Sigma }} = {\bf{\Sigma }}_{{\rm cov}}$ into (\ref{eqn:mse}), the \ac{mse} $E$ can be rewritten as
\begin{align}\label{eqn:proof3}
	\notag
	E & = {\rm{Tr}}\left( {{{\bf{\Sigma }}_{{\mathop{\rm cov}} }}} \right) \!-\!  {\rm{Tr}}\left( {{\bf{\Sigma }}_{{\mathop{\rm cov}} }}{\bf S} {{\left( { {\bf{S}}^{\rm{H}} {{\bf{\Sigma }}_{{\mathop{\rm cov}} }}{{\bf{S}}} \!+\! {\sigma ^2}{{\bf{I}}_{PM}}} \right)}^{ - 1}} {\bf{S}}^{\rm{H}}{{\bf{\Sigma }}_{{\mathop{\rm cov}} }} \right)
	\\
    & = {\rm Tr}\left({\bf \Sigma}_{{\bf h}|{\bf y}}\right){\big |}_{{\bf{\Sigma }} = {\bf{\Sigma }}_{{\rm cov}}}
\end{align}
which is exactly the trace of the posterior covariance ${\bf \Sigma}_{{\bf h}|{\bf y}}$ in (\ref{eqn:Sigma_O}) when ${\bf{\Sigma }} = {\bf{\Sigma }}_{{\rm cov}}$. In Bayesian linear regression, the posterior variance of sample path, i.e., each diagonal entry of ${\bf \Sigma}_{{\bf h}|{\bf y}}$, monotonically decreases as the number of samples increases \cite{oliver1990kriging}. Therefore, the \ac{mse} $E$ achieves its minimum when all port channels are fully observed (although it may be impractical), i.e., $PM=N$. By substituting $PM=N$ into (\ref{eqn:proof3}), we have 
\begin{align}
\notag
E_{\min}= & {\rm{Tr}}\left( {{{\bf{\Sigma }}_{{\mathop{\rm cov}} }}} \right) - {\rm{Tr}}\left( {{{\bf{\Sigma }}_{{\mathop{\rm cov}} }}{{\left( {{{\bf{\Sigma }}_{{\mathop{\rm cov}} }} + {\sigma ^2}{{\bf{I}}_N}} \right)}^{ - 1}}{{\bf{\Sigma }}_{{\mathop{\rm cov}} }}} \right)
\\ \notag \stackrel{(a)}{=} & {\rm Tr}\left(\left({\bf \Sigma}_{\rm cov}^{-1} + {\sigma^{-2}}{\bf I}\right)^{-1}\right)
\\ \stackrel{(b)}{=} & \sum\limits_{k=1}^K {\frac{1}{{\frac{1}{{{\lambda_k}}} + \frac{1}{{{\sigma ^2}}}}}} = \sum\limits_{k = 1}^K {\frac{{{\lambda _k}{\sigma ^2}}}{{{\lambda _k} + {\sigma ^2}}}} ,
\end{align}
where $(a)$ holds by using the well-known Sherman-Morrison-Woodbury formula ${({\bf{A}} + {\bf{BCD}})^{ - 1}} = {{\bf{A}}^{ - 1}} - {{\bf{A}}^{ - 1}}{\bf{B}}{\left( {{\bf{I}} + {\bf{CD}}{{\bf{A}}^{ - 1}}{\bf{B}}} \right)^{ - 1}}{\bf{CD}}{{\bf{A}}^{ - 1}}$ and $(b)$ holds according to the eigenvalue decomposition ${\bf \Sigma}_{\rm cov}={\bf U}{\rm diag}(\lambda_1,\cdots,\lambda_K){\bf U}^{\rm H}$ wherein ${\bf U}\in{\mathbb C}^{N\times K}$ is an orthogonal matrix. This completes the proof.


\footnotesize
\balance 
\bibliographystyle{IEEEtran}


\bibliography{IEEEabrv,reference}
	
\begin{IEEEbiography}[{\includegraphics[width=1in,height=1.25in,clip,keepaspectratio]{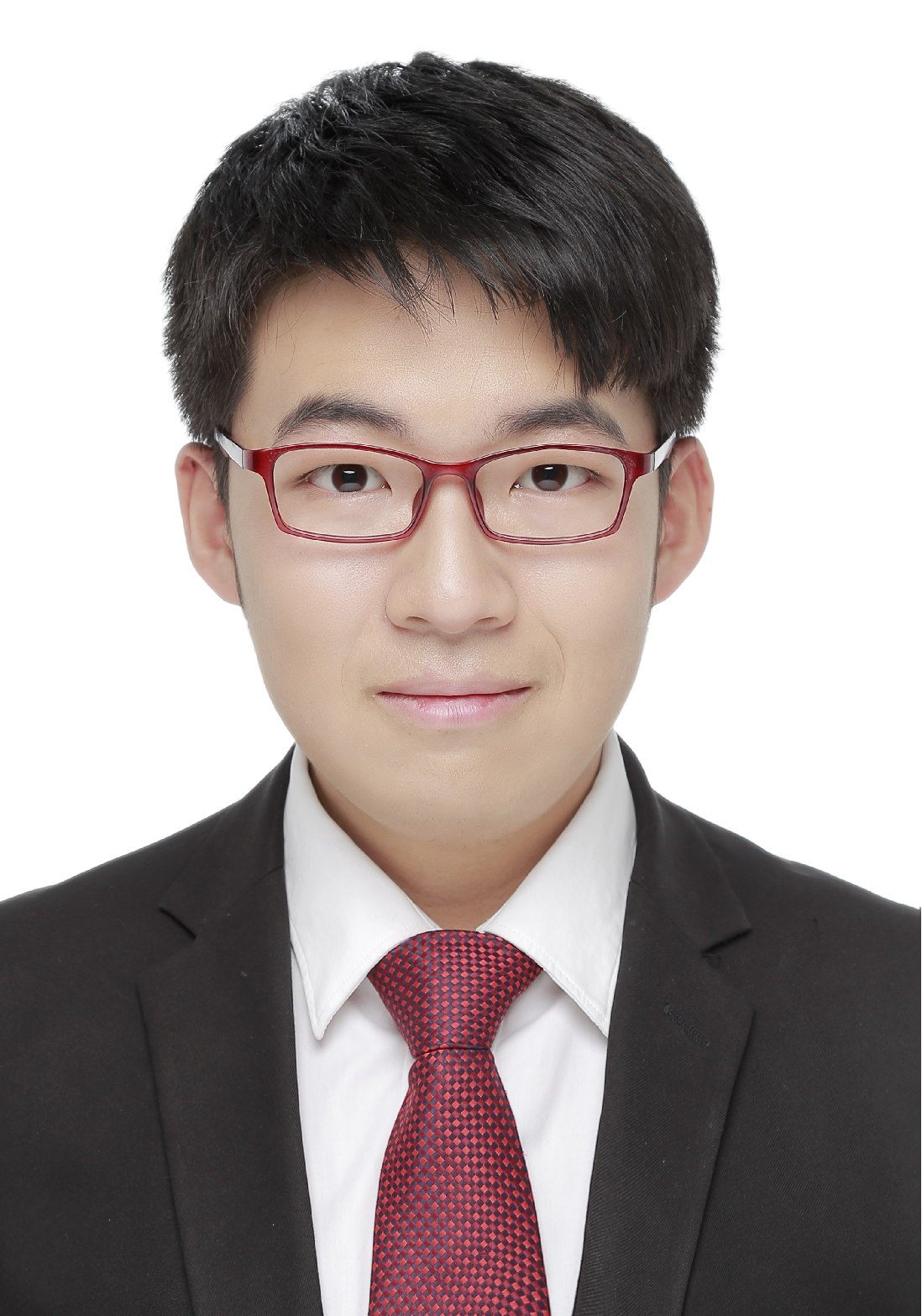}}]{Zijian Zhang}
	(Graduate Student Member, IEEE) received the B.E. degree in electronic engineering from Tsinghua University, Beijing, China, in 2020. He is currently working toward the Ph.D. degree in electronic engineering from Tsinghua University, Beijing, China.
	His research interests include massive MIMO, holographic MIMO (H-MIMO), reconfigurable intelligent surfaces (RISs), and fluid antenna systems (FASs). He is also an amateur in wireless localization and robotics. 
	
	He has authored several papers for the \textsc{IEEE Journal on Selected Areas in Communications}, the \textsc{IEEE Transactions on Signal Processing}, the \textsc{IEEE Transactions on Wireless Communications}, the \textsc{IEEE Transactions on Communications}, etc. He received the National Scholarship in 2019 and 2024. In 2024, he won the Special Scholarship of Tsinghua University, which annually awards 10 students out of over 40,000 graduate students in Tsinghua University. He was listed in Stanford University’s World’s Top 2\% Scientists in 2023.
\end{IEEEbiography}

\begin{IEEEbiography}[{\includegraphics[width=1in,height=1.25in,clip,keepaspectratio]{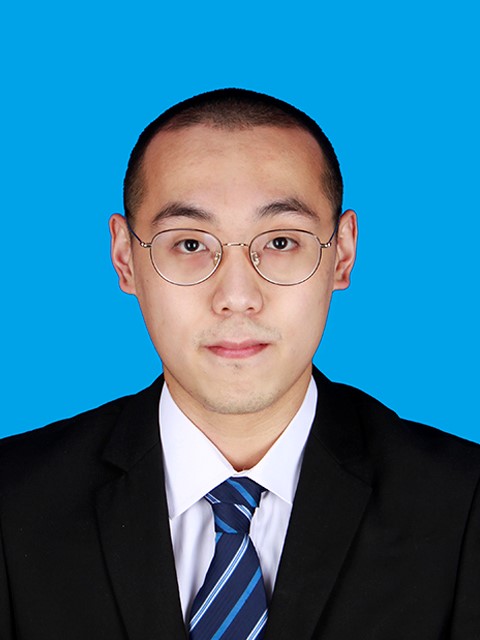}}]{Jieao Zhu} (Graduate Student Member, IEEE) received the B.E. degree in electronic engineering and the B.S. degree in applied mathematics from Tsinghua University, Beijing, China, in 2021, where he is currently pursuing the Ph.D. degree with the Department of Electronic Engineering. His research interests include electromagnetic information theory (EIT), coding theory, and quantum computing. He has received the National Scholarship in 2018 and 2020 and the Excellent Graduates of Beijing in 2021.
\end{IEEEbiography}

\begin{IEEEbiography}[{\includegraphics[width=1in,height=1.25in,clip,keepaspectratio]{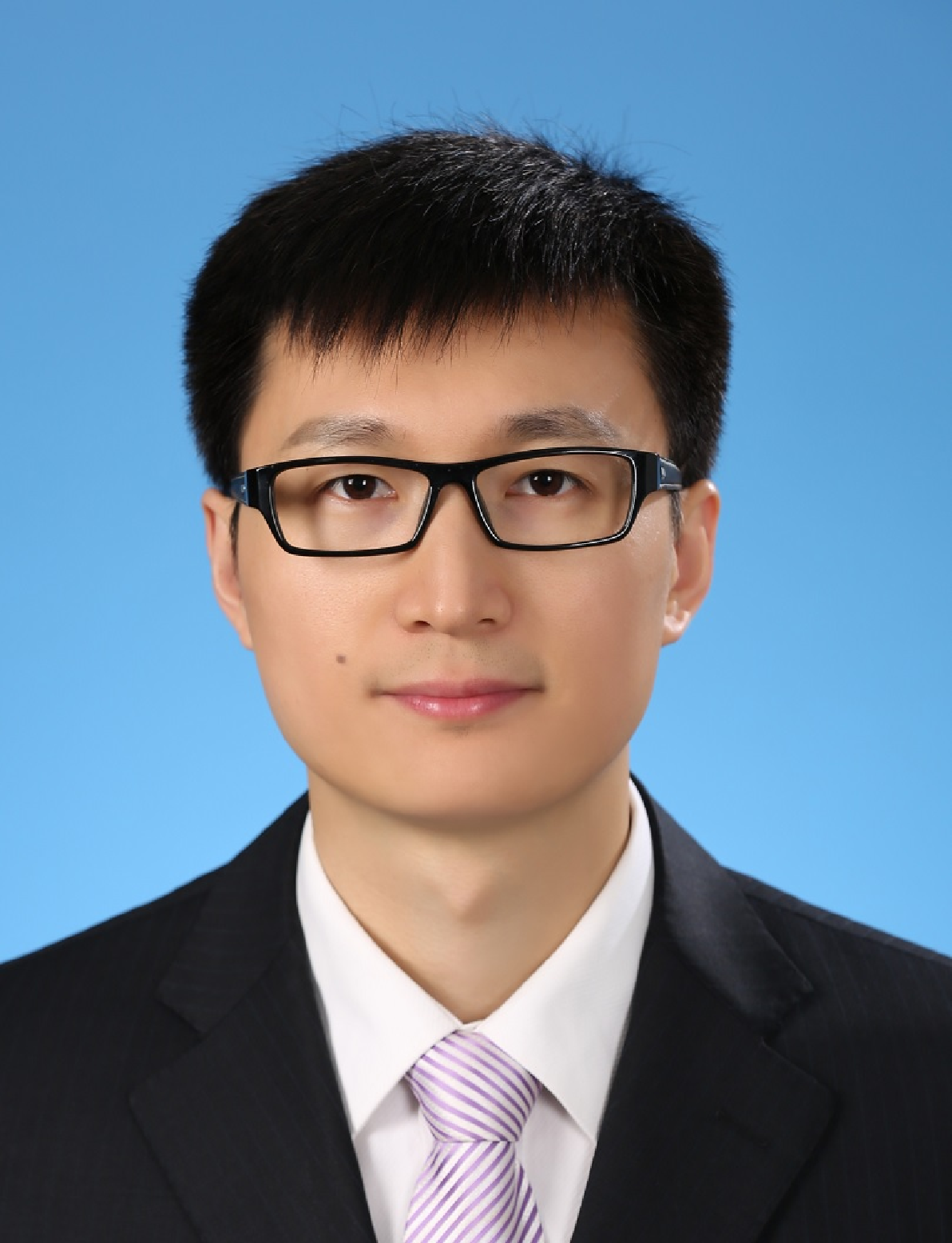}}]{Linglong Dai} (Fellow, IEEE) received the B.S. degree from Zhejiang University, Hangzhou, China, in 2003, the M.S. degree from the China Academy of Telecommunications Technology, Beijing, China, in 2006, and the Ph.D. degree from Tsinghua University, Beijing, in 2011. From 2011 to 2013, he was a Post-Doctoral Researcher with the Department of Electronic Engineering, Tsinghua University, where he was an Assistant Professor from 2013 to 2016, an Associate Professor from 2016 to 2022, and has been a Professor since 2022. His current research interests include massive MIMO, reconfigurable intelligent surface (RIS), millimeter-wave and Terahertz communications, near-field communications, machine learning for wireless communications, and electromagnetic information theory. 
	
He has coauthored the book {\it MmWave Massive MIMO: A Paradigm for 5G} (Academic Press, 2016). He has authored or coauthored over 100 IEEE journal papers and over 60 IEEE conference papers. He also holds over 20 granted patents. He has received five IEEE Best Paper Awards at the IEEE ICC 2013, the IEEE ICC 2014, the IEEE ICC 2017, the IEEE VTC 2017-Fall, the IEEE ICC 2018, and the IEEE GLOBECOM 2023. He has also received the Tsinghua University Outstanding Ph.D. Graduate Award in 2011, the Beijing Excellent Doctoral Dissertation Award in 2012, the China National Excellent Doctoral Dissertation Nomination Award in 2013, the URSI Young Scientist Award in 2014, the IEEE Transactions on Broadcasting Best Paper Award in 2015, the Electronics Letters Best Paper Award in 2016, the National Natural Science Foundation of China for Outstanding Young Scholars in 2017, the IEEE ComSoc Asia-Pacific Outstanding Young Researcher Award in 2017, the IEEE ComSoc Asia-Pacific Outstanding Paper Award in 2018, the China Communications Best Paper Award in 2019, the IEEE Access Best Multimedia Award in 2020, the IEEE Communications Society Leonard G. Abraham Prize in 2020, the IEEE ComSoc Stephen O. Rice Prize in 2022, the IEEE ICC Best Demo Award in 2022, and the National Science Foundation for Distinguished Young Scholars in 2023. He was listed as a Highly Cited Researcher by Clarivate Analytics from 2020 to 2024. He was elevated as an IEEE Fellow in 2022. 
\end{IEEEbiography}

\begin{IEEEbiography}[{\includegraphics[width=1in,height=1.25in,clip,keepaspectratio]{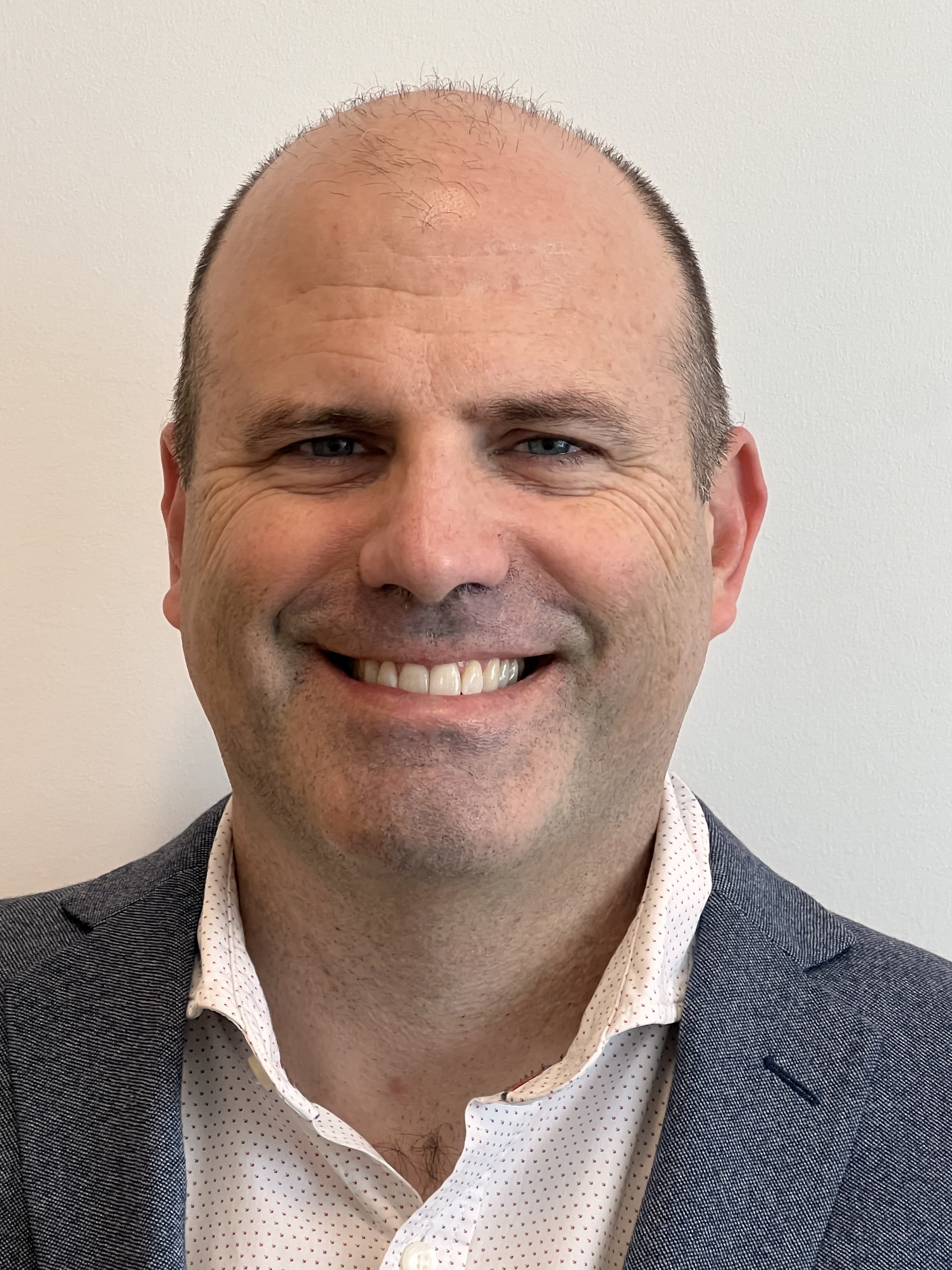}}]{Robert W. Heath Jr.} (Fellow, IEEE) is the Charles Lee Powell Chair in Wireless Communications in the Department of Electrical and Computer Engineering at the University of California, San Diego. He is also President and CEO of MIMO Wireless Inc. From 2020-2023 he was the Lampe Distinguished Professor at North Carolina State University and co-founder of 6GNC. From 2002-2020 he was with The University of Texas at Austin, most recently as Cockrell Family Regents Chair in Engineering and Director of UT SAVES.  He authored {\it Introduction to Wireless Digital Communication} (Prentice Hall, 2017) and {\it Digital Wireless Communication: Physical Layer Exploration Lab Using the NI USRP} (National Technology and Science Press, 2012), and co-authored {\it Millimeter Wave Wireless Communications} (Prentice Hall, 2014) and {\it Foundations of MIMO Communication} (Cambridge University Press, 2018). 

Dr. Heath has been a co-author of a number award winning conference and journal papers including recently the 2017 Marconi Prize Paper Award,  the 2019 IEEE Communications Society Stephen O. Rice Prize, the 2020 IEEE Signal Processing Society Donald G. Fink Overview Paper Award, the 2021 IEEE Vehicular Technology Society Neal Shepherd Memorial Best Propagation Paper Award, and the 2022 IEEE Vehicular Technology Society Best Vehicular Electronics Paper Award. Other notable awards include the 2017 EURASIP Technical Achievement award, the 2019 IEEE Kiyo Tomiyasu Award, the 2021 IEEE Vehicular Technology Society James Evans Avant Garde Award, and the 2025 IEEE/RSE James Clerk Maxwell Medal.  In 2017, he was selected as a Fellow of the National Academy of Inventors. In 2024, he was selected as a Fellow of the American Association for the Advancement of Science. He was a member-at-large on the IEEE Communications Society Board-of-Governors (2020-2022) and the IEEE Signal Processing Society Board-of-Governors (2016-2018). He was Editor-in-Chief of \textsc{IEEE Signal Processing Magazine} from 2018-2020. He is also a licensed Amateur Radio Operator, a Private Pilot, a registered Professional Engineer in Texas.
\end{IEEEbiography}	
\end{document}